\documentclass{aa}  
\usepackage{graphicx}
\usepackage{txfonts}
\usepackage{soul}
\begin{document}

   \title{Hydrogen Non-Equilibrium Ionisation Effects in Coronal Mass Ejections}

   \subtitle{ }

   \author{P. Pagano\inst{\ref{inst1}} \and
          A. Bemporad\inst{\ref{inst2}} \and
          D.H. Mackay\inst{\ref{inst1}}
          }

\authorrunning{Pagano et al.}
\titlerunning{Ionisation CMEs}

\institute{School of Mathematics and Statistics, University of St Andrews, North Haugh, St Andrews, Fife, Scotland KY16 9SS, UK \label{inst1}
            \and
            INAF -- Turin Astrophysical Observatory, via Osservatorio 20, 10025 Pino Torinese (TO), Italy \label{inst2}
 \\
      \email{pp25@st-andrews.ac.uk}
      }

   \date{ }

 
  \abstract
   {A new generation of coronagraphs to study the solar wind and Coronal Mass Ejections (CMEs) are being developed and launched. These coronagraphs will heavily rely on
   multi-channel observations where visible light (VL) and UV-EUV observations provide new plasma diagnostics.
   One of these instruments, Metis on board ESA-Solar Orbiter, will simultaneously observe VL and the UV Lyman-$\alpha$ line.
   The number of neutral Hydrogen atoms (a small fraction of coronal protons) is a key parameter for deriving plasma properties such as temperature from the observed Lyman-$\alpha$ line intensity.
   However, these measurements are significantly affected if non-equilibrium ionisation effects occur, which can be relevant during CMEs.} 
   {The aim of this work is to determine if non-equilibrium ionisation effects are relevant in CMEs and in particular
   when and in which regions of the CME plasma ionisation equilibrium can be assumed for data analysis.}
   {We use a magneto-hydrodynamic simulation of a magnetic flux rope ejection to generate a CME. From this we then reconstruct the ionisation state of Hydrogen atoms in the CME by evaluating both the advection of neutral and ionised Hydrogen atoms and the ionisation and recombination rates in the MHD simulation.}
   {We find that the equilibrium ionisation assumption holds mostly in the core of the CME, which is represented by a magnetic flux rope. In contrast
   non-equilibrium ionisation effects are significant at the CME front, where
   we find about 100 times more neutral Hydrogen atoms than prescribed by ionisation equilibrium conditions, even if this neutral Hydrogen excess might be difficult to identify due to projection effects.}
   {This work provides key information for the development of a new generation of diagnostic techniques that aim at combining visible light
   and Lyman-$\alpha$ line emissions. The results show that non-ionisation equilibrium effects need to be considered when we analyse CME fronts.
   To  incorrectly assume equilibrium ionisation in these regions would lead to a systematic underestimate of plasma temperatures.}


   \maketitle
%

\section{Introduction}

Coronal Mass Ejections (CMEs) are very dynamic and violent phenomena occurring in the solar corona.
Despite continuous observations and monitoring of these events
from ground and space
over the last decades, their elusive nature still presents a major challenge for solar physics. One of the main reasons is that measuring properties of the embedded plasma is always an arduous task. It is complicated by the optical thinness of the coronal plasma, the large speeds and rapid evolution of the CME, and the rapidly changing temperatures.
Due to this, remote sensing observations acquired at different wavelengths are combined and analysed with a number of diagnostic techniques
to derive the thermodynamic properties of the coronal plasma ejected in CMEs.

New diagnostic techniques currently under development
are based on the forthcoming remote sensing observations of the ultraviolet (UV) 
Lyman-$\alpha$ line emitted from neutral Hydrogen atoms. 
These diagnostics will be applied to
future data that will be acquired for instance by Metis \citep{2017SPIE10566E..0LA}, a new generation multi-channel coronagraph on board the ESA-Solar Orbiter mission
that will observe at the same time the solar corona in visible light (VL, polarised and non-polarised components) and in the Lyman-$\alpha$ line.
This will be done through the simultaneous acquisition of observations from a broad-band VL channel (580-640 nm)
and  a narrow-band one in the UV emission around 121.6 nm, where neutral Hydrogen atoms emit the Lyman-$\alpha$ line.

At the same time, other forthcoming missions will investigate the solar corona with UV observations centred on the Lyman-$\alpha$ line (the strongest line in the whole EUV coronal spectrum),
such as the LST instrument that will fly on-board the Chinese ASO-S mission in 2022 \citep{2016IAUS..320..436L}.
The key strength of these new instruments is the switch to multi-channel observations that
will allow the investigation of plasma properties that so far have been impossible to access due to the limitation of single channel VL observations.

These new observations will pose unprecedented challenges, because it will be essential 
to have the proper interpretative tools to obtain accurate derivations of CME properties.
For example, the simultaneous observation of VL (mostly photospheric emission scattered by free electrons in the solar corona)
and UV Lyman-$\alpha$ (emitted by neutral Hydrogen atoms mostly excited by chromospheric radiation) can open up a number of diagnostic possibilities. In a plasma, the intensity of the emissions coming from different ionisation stages of the same atomic species are 
connected to the abundance of each ionisation level, and can thus be a useful proxy for the plasma temperature. On the other hand, the Lyman-$\alpha$ emission from neutral H atoms can be used to derive the plasma temperatures by estimating the fraction of neutrals, once the plasma densities are known from VL data. 

These techniques were extensively discussed for instance by \citet{susino2016} who demonstrated, by using SOHO/LASCO and UVCS observations, that CME plasma temperatures can be derived by a direct combination of VL and UV Lyman-$\alpha$ intensities
even when neglecting spectroscopic information. Similar data analysis methods were also tested by \citet{Bemporad2018}, based on the analysis of synthetic data and under different assumptions mostly related with the integration along the line-of-sight (hereafter LOS) through the optically thin plasma. These techniques will likely provide the first ever 2D maps of plasma temperatures inside CMEs in the intermediate corona, and were also strengthened by the improved methods developed by \citet{BemporadPagano2015}
and \citet{Pagano2015b} to derive the column density of the CME plasma from 
polarised VL observations. Moreover, because of the Doppler dimming \citep{Noci1987}, the UV Lyman-$\alpha$ emission by coronal atoms also depends on their radial outflow speed, and thus the derivation of temperatures in CMEs also requires the technique recently developed by \citet{2019ApJ...880...41Y} to reconstruct the 2D distribution of plasma speed in the CME bodies from a sequence of VL images.

For all of the techniques described above to be accurate, we need to understand 
how the abundance of various ionisation levels evolves in dynamic events such as CMEs.
In this context, the key question is whether the ionisation equilibrium conditions
hold for Hydrogen atoms throughout the propagation of CMEs in the solar corona 
or is there instead a significant departure from ionisation equilibrium. The ionisation equilibrium hypothesis corresponds to the assumption that a balance exists between all processes causing the stripping of electrons from atoms of a
given element with all the recombination processes leading to lower ionisation stages.
For a long period, this simplifying hypothesis has been made in the analysis of UVCS 
observations of CMEs and post-CME current sheets in the intermediate corona \citep[see e.g.][]{Raymond2003a, Ciaravella2006}, where ionisation equilibrium was assumed to derive plasma temperatures.

The issue of the ion distribution in plasmas that are  involved in dynamic events
has puzzled solar physicists for decades, as non-equilibrium ionisation effects can become important in a number of scenarios.
\citet{2004A&A...416L..13K} and \citet{2019ApJ...871...18R} addressed the ionisation of the chromosphere during solar flares.
\citet{2019ApJ...879..111L} modelled the non-ionisation equilibrium effects following 
a heating event due to reconnection and found the locations where these effects can be relevant.
\citet{2019ApJ...872..123B} found that non-equilibrium ionisation effects are relevant during nanoflares.
\citet{2017ApJ...850...26S} studied the non-equilibrium ionisation in the solar wind.
\citet{2016ApJ...817...46M} described how the non-equilibrium ionisation affects the formation of IRIS lines. Recently, the ionisation equilibrium assumption was (often tacitly) 
performed by many authors to derive the 2D distributions of plasma temperatures during CMEs 
\citep[e.g.][]{hannah2013,dudik2014,su2016,aschwanden2017,frassati2019} from the analysis of EUV images acquired with the multiple filters provided by the AIA telescopes on-board SDO. 
In fact, the applied emission measure techniques
require the construction of synthetic EUV spectra for all lines emitted by different ions
in each instrument band-pass at various temperatures, to derive the instrument 
temperature responses. These synthetic spectra can be constructed so far
\citep[with numerical codes like the CHIANTI spectral code; see e.g.][and references therein]{dere2019} only by assuming ionisation equilibrium.

On the other hand, departures from ionisation equilibrium are expected to occur in many
dynamic phenomena in the solar atmosphere, as suggested 
by the significant spectral variability observed to occur on timescales shorter than the ion 
equilibration times. The principles of a numerical tool for the calculation of non-equilibrium 
ionisation states in the solar corona were illustrated by \citet{bradshaw2009}, and
calculations of the ionisation equilibrium timescales have been performed by
several authors, such as \citet{2010ApJ...718..583S} who derived the ionisation equilibrium timescales in the absence of flows. Recently \citet{dudik2017} reviewed the state of the art for these methods.
The departure from ionisation equilibrium in CME shocks for Oxygen and Silicon was justified with a numerical model by \citet{Pagano2008}.
The violation of the 
ionisation equilibrium assumption in the post-CME EUV dimming region was 
reported by \citet{imada2011} in the higher temperature range. Later
\citet{shen2013} showed that, because of non-equilibrium ionisation effects, 
the analysis of EUV imaging observations of post-CME current sheets would lead 
to an under (over) estimate of plasma temperatures at low (large) 
heights by about a factor of two if equilibrium is assumed. On the other hand,
\citet{kocher2018} reported that in a CME core, departures from ionisation 
equilibrium were within 10\% for Hydrogen and Helium atoms in the erupting filament.
Very recently, possible observational consequences of non-equilibrium ionisation
in the EUV imaging of CME related phenomena (such as post-flare arcades and CME-driven shocks)
were investigated by \cite{lee2019}: the authors re-analysed previous dynamic events
studied with an emission measure technique applied to EUV and soft X-ray images and 
concluded that the temperature of hotter plasma not in ionisation equilibrium is 
lower than the temperature calculated by assuming ionisation equilibrium. These
effects can be important not only for the study of CMEs, but also for elemental
abundance determination in stationary solar wind flows \citep{shi2019}.

Similar analyses have often been performed for observations of erupting 
events in the transition region and in the inner corona (R < 1.5 R$_\odot$),
but not very often for CMEs observed higher up in the intermediate corona during the expansion phase (R > 1.5 R$_\odot$). At these altitudes the problem of ionisation equilibrium
was discussed for instance by \citet{akmal2001}, dealing with UVCS observations of a CME 
and the reconstruction of the thermal evolution of the observed plasma. More recently \citet{landi2010} and \citet{murphy2011} have reported similar observations of CMEs with UVCS 
spectra and other EUV imagers 
and spectrometers. In the interpretation of UVCS observations the problem was investigated by \citet{Ciaravella2001} also with numerical simulations, inferring the plasma heating rate in a CME by assuming ionisation equilibrium. Later, \citet{Pagano2008} simulated the UV spectroscopic emission from a CME (focusing on the evolution during a shock transit) as observed by UVCS 
by including also the effects of non-equilibrium ionisation. More recently, a non-equilibrium ionisation analysis
was performed by \citet{jejcic2017} to derive the physical parameters of an erupting prominence
observed spectroscopically with UVCS. While a significant amount of work has been carried out, none of these works focused on the 
evolution of neutral Hydrogen ionisation states for the analysis of the Lyman-$\alpha$ line
intensity evolution observed during CMEs.

Hence in summary, the interpretation of coronagraphic observations of CMEs in the Lyman-$\alpha$ line that will be provided by forthcoming instruments such as Metis on-board Solar Orbiter and LST on-board ASO-S missions will require a deep understanding of which regions in CME bodies can be considered in ionisation equilibrium.
In order to answer this question, in this paper we adopt a theoretical approach by using our realistic numerical modelling of the propagation of a CME to study the evolution of the ionisation state of
Hydrogen atoms during the CME. Through this we can relate our results with the ionisation state
that is prescribed by ionisation equilibrium conditions.

To reach our goals we start from the MHD simulation of \citet{Pagano2014} of a magnetic flux rope ejection.
Magnetic flux ropes are structures found in the solar corona
that usually lie on polarity inversion lines \citep[PILs][]{Cheng2010}
that are known to suddenly erupt and produce CMEs \citep{Chen2011,Yan2017, Song2014, HowardDeForest2014}. 
We first use a magnetofrictional model to describe the formation phase of magnetic flux rope using a quasi-static and magnetically dominated approach
and we switch to MHD simulations when the magnetic configuration is unstable.
In the MHD simulations, we solve the general form of the MHD equations
that are appropriate to describe a dynamic evolution where the dominant forces can be magnetic, pressure gradients, or gravity. 

Once we have solved the MHD equations for the eruption of the magnetic flux rope, 
we devise a technique to reconstruct the history of the ionisation state of Hydrogen atoms in the 3D domain of the MHD equations as a function of time.
Such an approach has already been successfully adopted by \citet{Pagano2008}
for the study of the emission of Oxygen and Silicon lines in shocks connected with CMEs.
In order to reconstruct the ionisation state of Hydrogen in our MHD domain,
we separately estimate the advection of neutral and ionised Hydrogen atoms
and the local rate of ionisation and recombination processes. Finally we combine
these two terms to obtain the abundances of neutral and ionised Hydrogen atoms 
as a function of space and time.

The paper is structured as follows.
In Sec.\ref{mhsimulation} we summarise the key properties of the MHD simulation 
we use here and in Sec.\ref{methodionisation} we describe the technique we use to reconstruct the history of the ionisation state of Hydrogen atoms. Next
in Sec.\ref{results} we illustrate the results of our study and we draw our conclusions in Sec.\ref{conclusions}.

\section{MHD simulation}
\label{mhsimulation}

In the present paper, we analyse 
the ionisation state of Hydrogen atoms during a CME
starting from the data
obtained in the MHD simulation of \citet{Pagano2014}.
In \citet{Pagano2014} the simulation of magnetic flux rope ejection is carried out by coupling the magnetofrictional model of \citet{MackayVanBallegooijen2006A,Mackay2011} with a magnetohydrodynamic (MHD) simulation.
This technique, initiated by \citet{Pagano2013a} and used in a number of studies \citep{Pagano2013b, Pagano2014, Pagano2015a, Pagano2015b, Rodkin2017, Pagano2018}, has proven to be effective in modelling the full life span of magnetic flux ropes.

In that simulation, we used the MPI-AMRVAC software~\citep{Porth2014} to solve the MHD equations,
where solar gravity, anisotropic thermal conduction, and optically thin radiative losses are treated as source terms:
\begin{equation}
\label{mass}
\frac{\partial\rho}{\partial t}+\vec{\nabla}\cdot(\rho\vec{v})=0,
\end{equation}
\begin{equation}
\label{momentum}
\frac{\partial\rho\vec{v}}{\partial t}+\vec{\nabla}\cdot(\rho\vec{v}\vec{v})
   +\nabla p-\frac{(\vec{\nabla}\times\vec{B})\times\vec{B})}{4\pi}=+\rho\vec{g},
\end{equation}
\begin{equation}
\label{induction}
\frac{\partial\vec{B}}{\partial t}-\vec{\nabla}\times(\vec{v}\times\vec{B})=0,
\end{equation}
\begin{equation}
\label{energy}
\frac{\partial e}{\partial t}+\vec{\nabla}\cdot[(e+p)\vec{v}]=\rho\vec{g}\cdot\vec{v}-n^2\chi(T)-\nabla\cdot\vec{F_c},
\end{equation}
where $t$ is the time, $\rho$ the density,
$\vec{v}$ velocity, $p$ thermal pressure, $\vec{B}$ magnetic field,
$e$ the total energy,
$n$ number density, $\vec{F_c}$ the conductive flux according to \citet{Spitzer1962},
and $\chi(T)$ the radiative losses per unit emission measure \citep{Colgan2008}.
To close the set of Eqs. \ref{mass}-\ref{energy} we have a relation between internal, total, kinetic, and magnetic energy
\begin{equation}
\label{enercouple}
\frac{p}{\gamma-1}=e-\frac{1}{2}\rho\vec{v}^2-\frac{\vec{B}^2}{8\pi},
\end{equation}
where $\gamma=5/3$ denotes the ratio of specific heat, and
the expression for solar gravitational acceleration is
\begin{equation}
\label{solargravity}
\vec{g}=-\frac{G M_{\odot}}{r^2}\hat{r},
\end{equation}
where $G$ is the gravitational constant, $M_{\odot}$ denotes the mass of the Sun,
$r$ is the radial distance from the center of the Sun
and $\hat{r}$ is the corresponding unit vector.
The spatial domain of this simulation extends over $3$ $R_{\odot}$ in the radial direction starting from
$r=R_{\odot}$. The colatitude, $\theta$, spans from $\theta=30^{\circ}$ to
$\theta=100^{\circ}$ and the longitude, $\phi$, spans over $90^{\circ}$.
For more details,
we refer the reader to a series of works:
\citet{MackayVanBallegooijen2006A}
where the initial pre-eruptive configuration is derived using a magnetofrictional model,
\citet{Pagano2013a} where for the first time we coupled
the eruptive initial conditions with the MHD simulations,
\citet{Pagano2013b} where we study the effect 
of gravity on the propagation of CMEs in the corona, 
and finally \citet{Pagano2014} where
we included the effects of non-ideal MHD.

In the magnetofrictional model two neighbouring magnetic bipoles are allowed to 
evolve 
under the effect of differential rotation, meridional flows and surface diffusion.
As one of the two bipoles is initially more sheared than the other
a magnetic flux rope forms in the corona above its polarity inversion line (PIL) after 19 days of evolution \citep{MackayVanBallegooijen2006A}.
At the same time, due to the magnetic field configuration another PIL separates the two bipoles 
and above this one a null point is present.  With evolution and the formation of a flux rope the magnetic field configuration obtained is not stable and a significant Lorentz force is present underneath the magnetic flux rope.
In this MHD simulation, we also construct an atmosphere around the magnetic field configuration
where the magnetic flux rope is modelled initially to be colder and denser 
than its surroundings.
This is done by imposing a temperature distribution dependent of the $\theta$-component of the magnetic field, so to have cold plasma where the we have a magnetic flux rope that develops with an $\theta$-directed axial magnetic field in this configuration and, at the same time, maintaining an horizontal thermal pressure balance,
so that these cold regions are also denser than their surroundings. 
In the initial condition of this simulation the minimum temperature is set to $T_{min}=10^4$ in a corona at $T=2$ $MK$, leading to a magnetic flux rope structure that is up to 200 times denser than the background corona.
More details on how this atmosphere is constructed can be found in \citet{Pagano2014}.

\begin{figure}[!htb]
\centering
\includegraphics[scale=0.20]{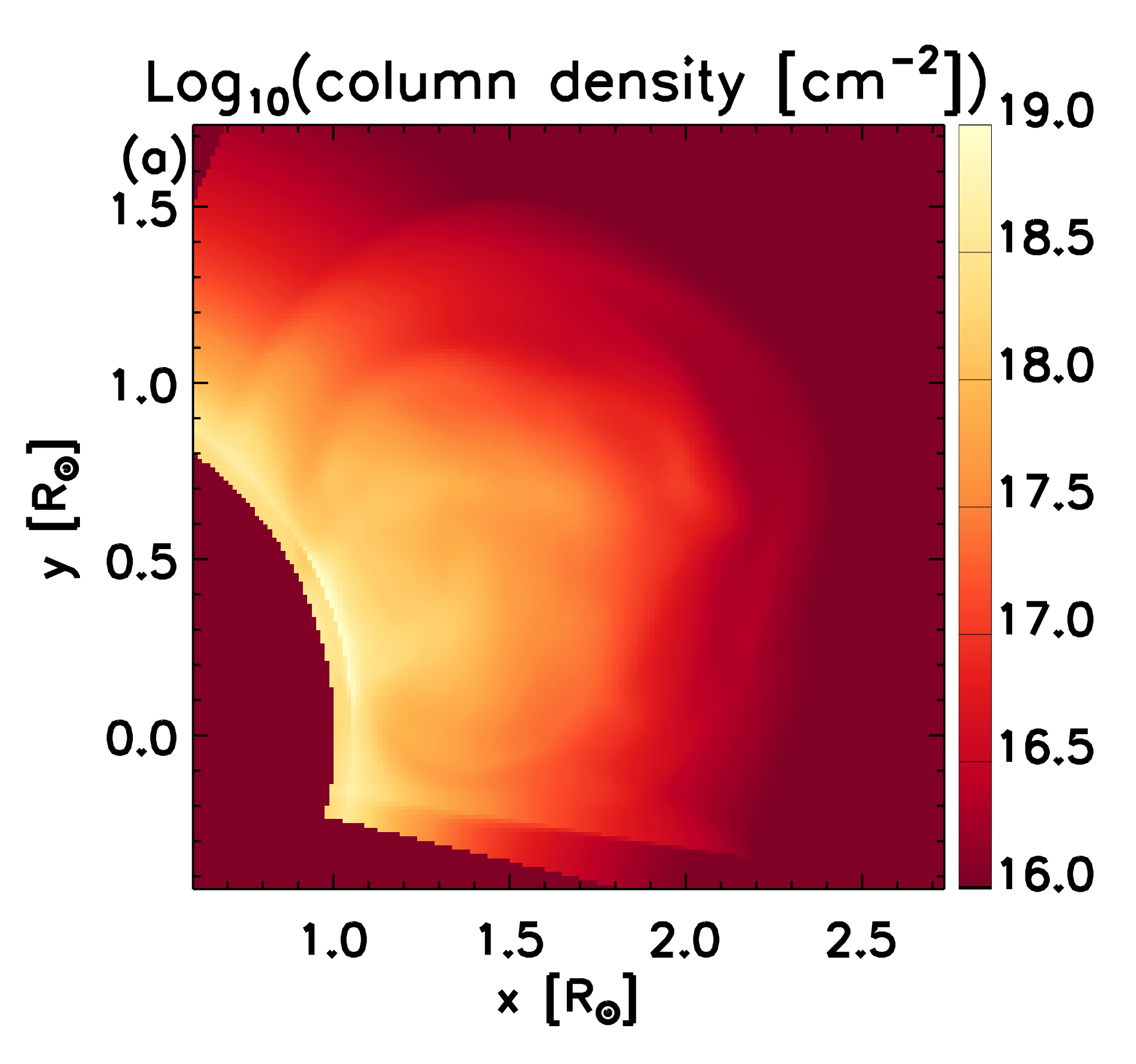}
\includegraphics[scale=0.20]{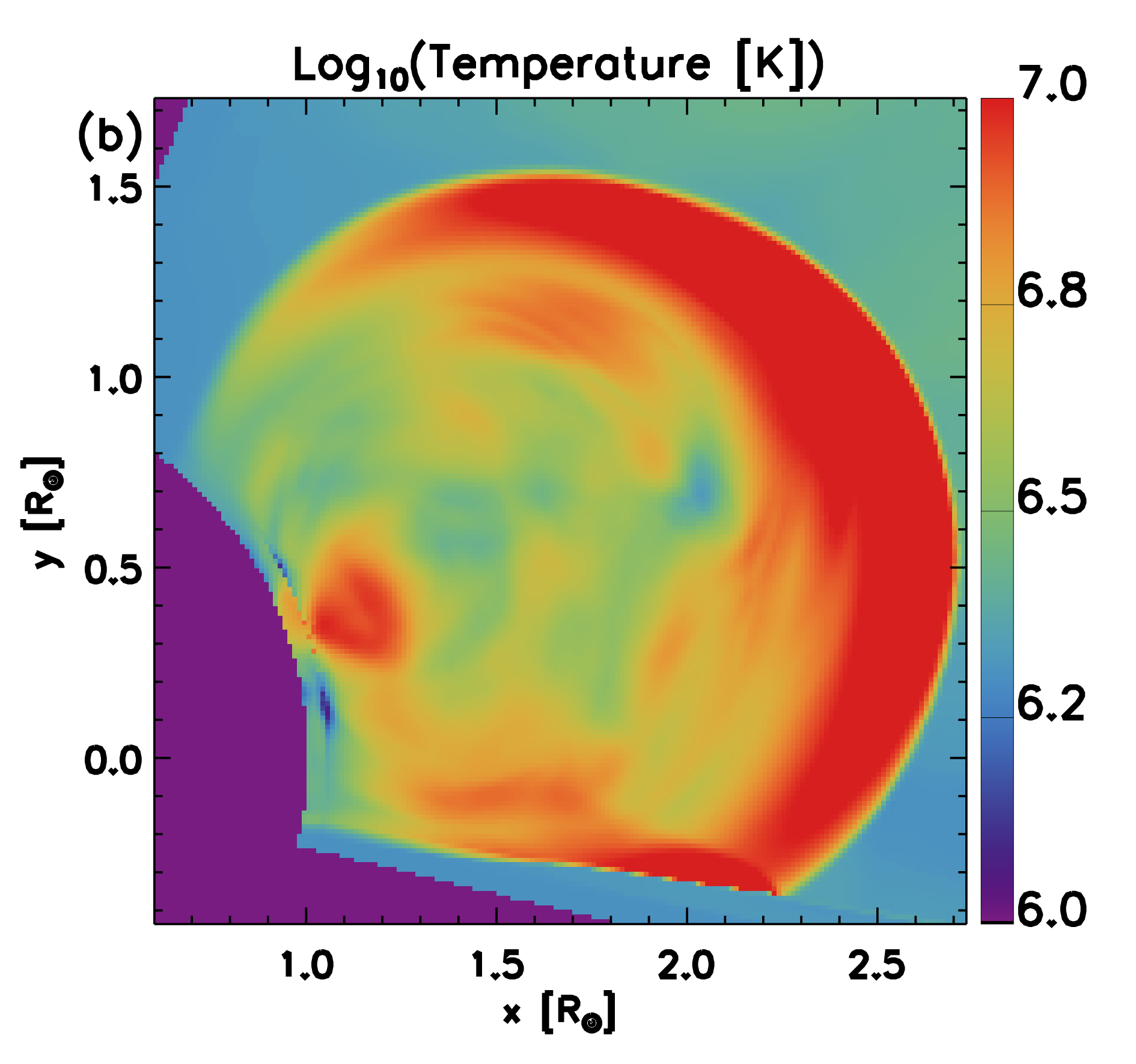}
\caption{(a) Map of column density and
(b) map of temperature averaged over plasma density
of the simulation at $t=19.7$ $minutes$ interpolated in a cartesian frame.
We only show the field of view over $1$ $R_{\odot}$.}
\label{MHDneletemp}
\end{figure}

The flux rope is then ejected out of the corona due to
the Lorentz force distribution that builds-up over 19 days of evolution.
Figure \ref{MHDneletemp} shows the simulation after $19.7$ minutes of evolution for both the column density (Fig \ref{MHDneletemp}a) and the density-weighted temperature along the LOS, $T_{\mathrm{LOS}}$ (Fig \ref{MHDneletemp}b)
\begin{equation}
\label{averaget}
T_{\mathrm{LOS}}=\frac{\int_{-\infty}^{+\infty} \rho T \, dz}
{\int_{-\infty}^{+\infty} \rho \, dz}
.\end{equation}

The key features of this evolution are that we find plasma travelling outwards
because of the ejection and the plasma distribution closely resembles a 
three-component CME,
as a dense front is followed by a denser core with a void in between.
A similar pattern is found in the temperature map, 
where we find a hot front of the ejection due to the compression where
a colder region lags behind.

\begin{figure*}
\centering
\includegraphics[scale=0.60]{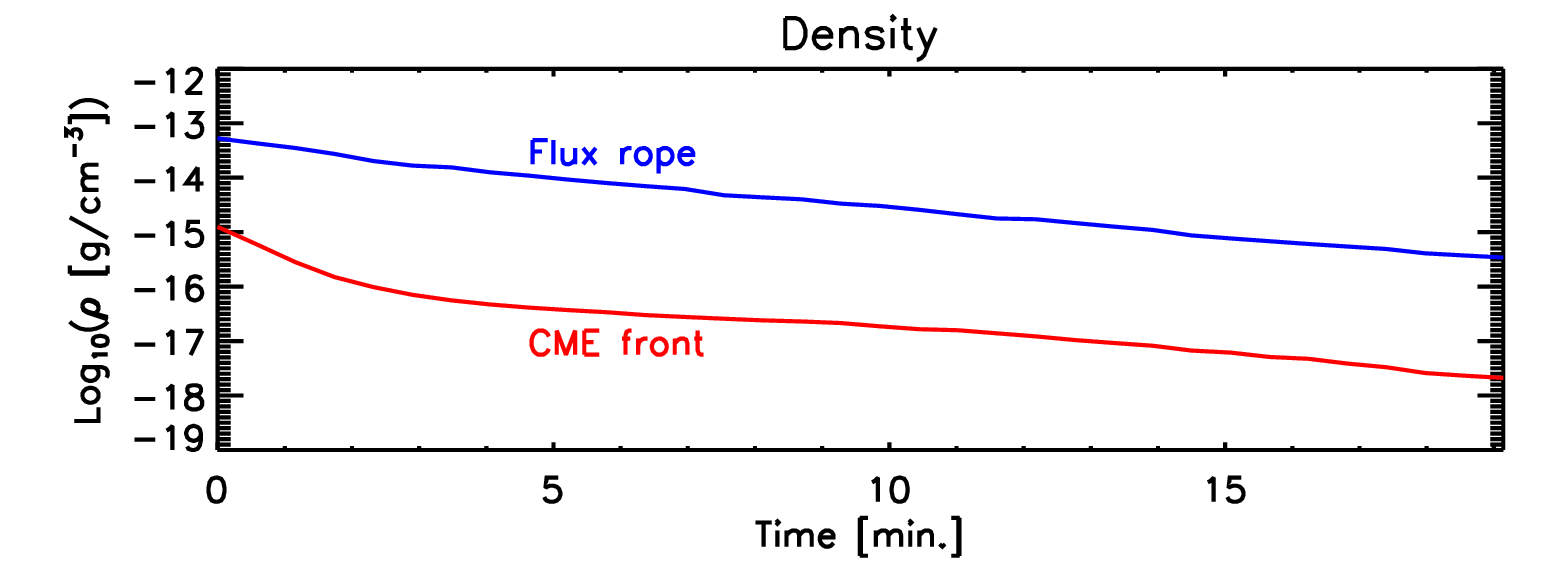}

\includegraphics[scale=0.60]{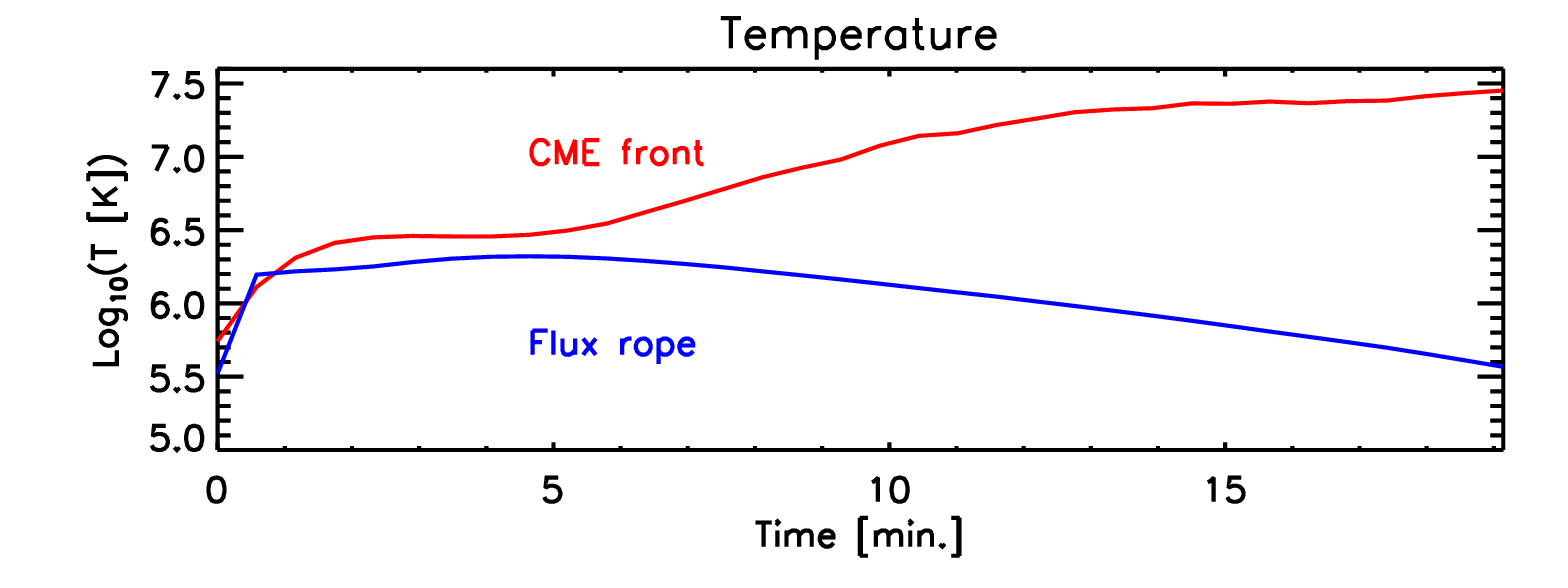}
\caption{Time evolution of plasma density (top panel) and temperature (bottom panel) 
at the centre of the flux rope (blue line) and at the CME front (red line).}
\label{frontfrtemprho}
\end{figure*}

One important aspect of the simulation is that the direction along which the magnetic flux rope expands is not radial, and instead the magnetic flux rope is ejected in the direction of the null point
that lies between the two bipoles.
Therefore, knowing the direction of propagation and the location of the magnetic flux rope and the CME front on this direction as function of time, we can
follow the plasma density and temperature inside these structures.
Fig.\ref{frontfrtemprho} shows the density and temperature at the centre of the flux rope and at the CME front as a function of time.

We find that the densities of both the centre of the flux rope and the CME front decrease in time
as all structures expand and move into regions of lower density higher in the solar corona.
The evolution of the plasma temperature is instead more complex.
As explained in \citet{Pagano2014}, the temperature of the magnetic flux rope immediately increases during
the very early stage of the MHD simulation due to the conversion of magnetic energy into thermal energy.
Later on, as the magnetic flux rope expands adiabatically, it also slowly cools down.
The CME front shows instead a constant temperature increase.
It basically coincides with the magnetic flux rope at the beginning of the simulation ($t<6$ $min$)
as the two structures are very close and it is not possible to identify a very distinct CME front.
Afterwards the compression occurring at the CME front heats the plasma to over $10$ $MK$,
where it more or less plateaus till the end of the simulation.
While a temperature significantly above $10$ $MK$ may seem unrealistic in the solar corona, such values
are nevertheless in the correct order of magnitude 
for the tenuous plasma of the outer corona violently heated by the compression generated from a violent CME.

\section{Ionisation state of plasma}
\label{methodionisation}

In this section we illustrate how we reconstruct the ionisation state of the Hydrogen atoms in the plasma 
as a function of space and time in our MHD simulation.
This is done by post-processing the results of the MHD simulation, 
thus assuming that the ionisation state of Hydrogen does not affect
the dynamic and thermal evolution of the CME in the MHD simulation.
This assumption is based on the fact that
the amount of neutral Hydrogen atoms is usually many orders of magnitude smaller 
than coronal protons and electrons, and in collisionless plasma even the very few 
neutral atoms are not interacting with other particles neither by collisions, nor by
magnetic fields. This strategy has been already been adopted in \citet{Pagano2008} 
for the derivation of the ionisation state of Oxygen and Silicon.

In order to reconstruct the ionisation state of Hydrogen we need to solve the following coupled differential equations
\begin{equation}
\label{timenhi}
\frac{\partial n_{HI}}{\partial t}=-\vec{\nabla}\cdot(n_{HI}\vec{v})+n_{HII}\left(\alpha_{HII}n_{HII}-q_{HI}n_{HI}\right)
\end{equation}
\begin{equation}
\label{timenhii}
\frac{\partial n_{HII}}{\partial t}=-\vec{\nabla}\cdot(n_{HII}\vec{v})+n_{HII}\left(q_{HI}n_{HI}-\alpha_{HII}n_{HII}\right)
\end{equation}
where $n_{HI}$ and $n_{HII}$ are the numbers densities of neutral and ionised Hydrogen atoms respectively as function of space and time,
$\alpha_{HII}$ is the recombination rate by collisions of ionised Hydrogen,
$q_{HI}$ is the ionisation rate by collisions of neutral Hydrogen,
$t$ is the independent variable time and $\vec{v}$ is the velocity field that is input from the MHD simulation.
As we have the velocity field only at the output cadence of the MHD simulation, that is $34.8$ second,
we linearly interpolate for intermediate times.
The values of $\alpha_{HII}$ and $q_{HI}$ are taken from CHIANTI spectral code \citep[][]{dere2019} as a function of the temperature $T$ of the plasma
and number density $n_e$ of the electrons.

The general solution of these equations provides the temporal and spacial dependent abundance of 
$n_{HI}$ and $n_{HII}$, including solutions that describe the 
out of ionisation equilibrium configurations.
In order to solve these equations, we assume an initial condition at $t=0$
where the Hydrogen atoms are in ionisation equilibrium.
\begin{equation}
n_{HII}=\rho/m_p
\label{nhiieq}
\end{equation}
\begin{equation}
\left[\alpha_{HII}n_{HII}-q_{HI}n_{HI}\right]=0.
\label{nhieq}
\end{equation}
Eq.\ref{nhiieq} expresses that the number of ionised Hydrogen atoms is the same as the number density of electrons as described by the MHD equations,
and in Eq.\ref{nhieq} we derive the number of neutral Hydrogen atoms from the number of ionised Hydrogen atoms when the number of recombinations 
matches the number of ionisations.
As the temperature is not uniform at $t=0$ in our MHD simulation, neither is the ratio $n_{HI}/n_{HII}$.

\subsection{Solution of the ionisation state of Hydrogen}

In order to solve the two differential equations (Eq.\ref{timenhi}-\ref{timenhii}) that govern the 
ionisation state of Hydrogen we need to evaluate 
the left hand side of both equations to obtain the rate of change
for $n_{HII}$ and $n_{HI}$ as a function of space at a given time $t$.
To do so, we assume that we can split the right hand side terms of 
Eq.\ref{timenhi} and Eq.\ref{timenhii}
into two terms, the advection term and the ionisation/recombination term,
and that accordingly we can integrate Eq.\ref{timenhi} and Eq.\ref{timenhii} in time
to obtain the finite variation of $n_{HII}$ and $n_{HI}$ over a finite integration time:
\begin{equation}
\label{timenhisplit}
\Delta n_{HI}=\left[ADV_{HI}\right]+\left[IR_{HI}\right]
\end{equation}
\begin{equation}
\label{timenhiisplit}
\Delta n_{HII}=\left[ADV_{HII}\right]+\left[IR_{HII}\right].
\end{equation}
We then separately evaluate each term on the right hand side of Eq.\ref{timenhisplit} and Eq.\ref{timenhiisplit}
and add their contributions to evaluate the left hand side.
In the next two sections we describe how we evaluate the advection terms and the ionisation/recombination terms.

\subsubsection{Advection term}
The advection term describes the motion of the Hydrogen atoms (neutrals and ions) 
due to the velocity field derived from the solution of the MHD equations.
For neutral Hydrogen atoms this term can be written as:
\begin{equation}
\label{advectionterm}
\left[ADV_{HI}\right]=-\vec{\nabla}\cdot(n_{HI}\vec{v}) \Delta t
\end{equation}
where $\Delta t$ is the finite integration time step.
In order to estimate $\left[ADV_{HI}\right]$ and $\left[ADV_{HII}\right]$ we use an explicit Godunov scheme \citep{Godunov1959}
where we use boundary conditions consistent with the ones adopted in solving 
the MHD equations.
In the Godunov scheme we need to estimate the flux of ionised $\vec{F^{n_{HII}}}$ and neutral $\vec{F^{n_{HI}}}$ Hydrogen atoms  that cross each
face of the finite difference cells of the MHD simulation grid. 
In order for this explicit scheme to work, we need to limit
the integration time step, $\Delta t$, so that the change of the number of neutral or ionised Hydrogen atoms
due to the fluxes is smaller than the number of atoms in each cell.
For instance, considering the $r$ direction and the number density $n_{HII}$,
this constraint reads:
\begin{equation}
\label{deltatime}
\Delta t< c_{CFL}\frac{\Delta r}{F_r^{n_{HII}}}.
\end{equation}
where $c_{CFL}<1$ is a parameter to satisfy the CFL (Courant-Friedrichs-Lewy) 
condition \citep{1928MatAn.100...32C}. It is
chosen to be $c_{CFL}=0.75$ in this work.
We then take the minimum from all the cells, for all directions and for $n_{HII}$ and $n_{HI}$ to estimate the integration time $\Delta t$.
However, because of the high heterogeneity in the initial $n_{HII}$ and $n_{HI}$ distributions
and the steep dependence of $\alpha_{HII}$ and $q_{HI}$ on the temperature,
the explicit solution of the advection term can lead to values of $\Delta t$
too small for reasonable computational times.

In order to circumvent this problem, we apply two measures.
First of all we use a smoothed distribution of $\rho$, $T$, and $\vec{v}$
as input from the MHD equations, and second we apply a flux limiter for the Godunov scheme.
The smoothing simply consists of replacing the values of each quantity at time $t$ in each computational cell
with the average among a volume of $3\times3\times3$ cells centred around the same cell.
As far as the flux limiter is concerned, we first estimate the $\Delta t$ resulting from
a given distribution of $n_{HII}$, $n_{HI}$, and $\vec{v}$.
If $\Delta t$ is smaller than $0.1$ $s$ in a cell, we redistribute a fraction of $10^{-4}$ of the 
number of ionised and neutral Hydrogen atoms from that cell to the neighbouring cells.
This is iterated until $c_{CFL}\Delta t\ge0.1$ $s$.
When this condition is satisfied we pick $\Delta t=0.1$ $s$ that is certainly smaller than the $\Delta t$ required by Eq.\ref{deltatime}.
This flux limiter is equivalent to a diffusion term that operates on a faster time scale of the advection of the bulk plasma motion,
such as turbulence.

\subsubsection{Ionisation/Recombination term}
The ionisation/recombination term expresses the rate at which
neutral or ionised Hydrogen atoms change ionisation state
per unit of time.
As Hydrogen atoms have only two possible ionised state we can write that:
\begin{equation}
\label{ionrecterm}
\left[IR_{HI}\right]=-\left[IR_{HII}\right]=n_{HII}\left(\alpha_{HII}n_{HII}-q_{HI}n_{HI}\right)\Delta t.
\end{equation}
In order to estimate the value of $\left[IR_{HI}\right]$ and $\left[IR_{HII}\right]$ 
we use an implicit scheme.
In this implicit scheme we set $\Delta t=0.1$, as we do for the advection term, to ensure
that the two terms are integrated over the same time interval and 
then we solve the implicit scheme associated with the finite difference expression for Eq.\ref{ionrecterm} for $\left[IR_{HI}\right]$ and the corresponding equation for  $\left[IR_{HII}\right]$.

This corresponds to solving the system of quadratic equations
\begin{equation}
\label{nhiquad}
n_{HI}^{t+\Delta t}-n_{HI}^{t}  =n_{HII}^{t+\Delta t}\left(n_{HII}^{t+\Delta t}\alpha_{HII}-n_{HI}^{t+\Delta t}q_{HI}\right)\Delta t
\end{equation}
\begin{equation}
\label{nhiiquad}
n_{HII}^{t+\Delta t}-n_{HII}^{t}=n_{HII}^{t+\Delta t}\left(n_{HI}^{t+\Delta t}q_{HI}-n_{HII}^{t+\Delta t}\alpha_{HII}\right)\Delta t
\end{equation}
for $n_{HI}^{t+\Delta t}$ and $n_{HII}^{t+\Delta t}$ that are the values for $n_{HI}$ and $n_{HII}$ at the time $t+\Delta t$,
whereas $n_{HI}^{t}$ and $n_{HII}^{t}$ are the known values of $n_{HI}$ and $n_{HII}$ at the time $t$.

In a few cells near the lower boundary of the MHD simulation where boundary conditions 
lead to sharp gradients in the number density of ionised and neutral Hydrogen atoms
or rapid changes in the plasma temperature occur,
it is possible that $n_{HII}$ or $n_{HI}$ could become negative.
For this reason we always impose ionisation equilibrium over 
the first 10 layers of the computational domain in $r$,
till $r<1.17$ $R_{\odot}$,
and when this happens we 
impose equilibrium ionisation also on other cells to restore a physically meaningful solution.
It should be noted that as soon as the magnetic flux rope is ejected
from its initial position near the lower boundary at $t=0$,
these locations are no longer relevant for the analysis of the CME ionisation state as the magnetic flux rope is quickly located far from the lower boundary where these issues arise.

\section{Results}
\label{results}

Having solved the MHD equations for a magnetic flux rope ejection
and then reconstructed the history of the ionisation state of Hydrogen atoms
we now discuss how the abundances of ionised and neutral Hydrogen atoms without
ionisation equilibrium compare to those derived in the usual hypothesis of 
ionisation equilibrium.

\begin{figure}
\centering
\includegraphics[scale=0.13]{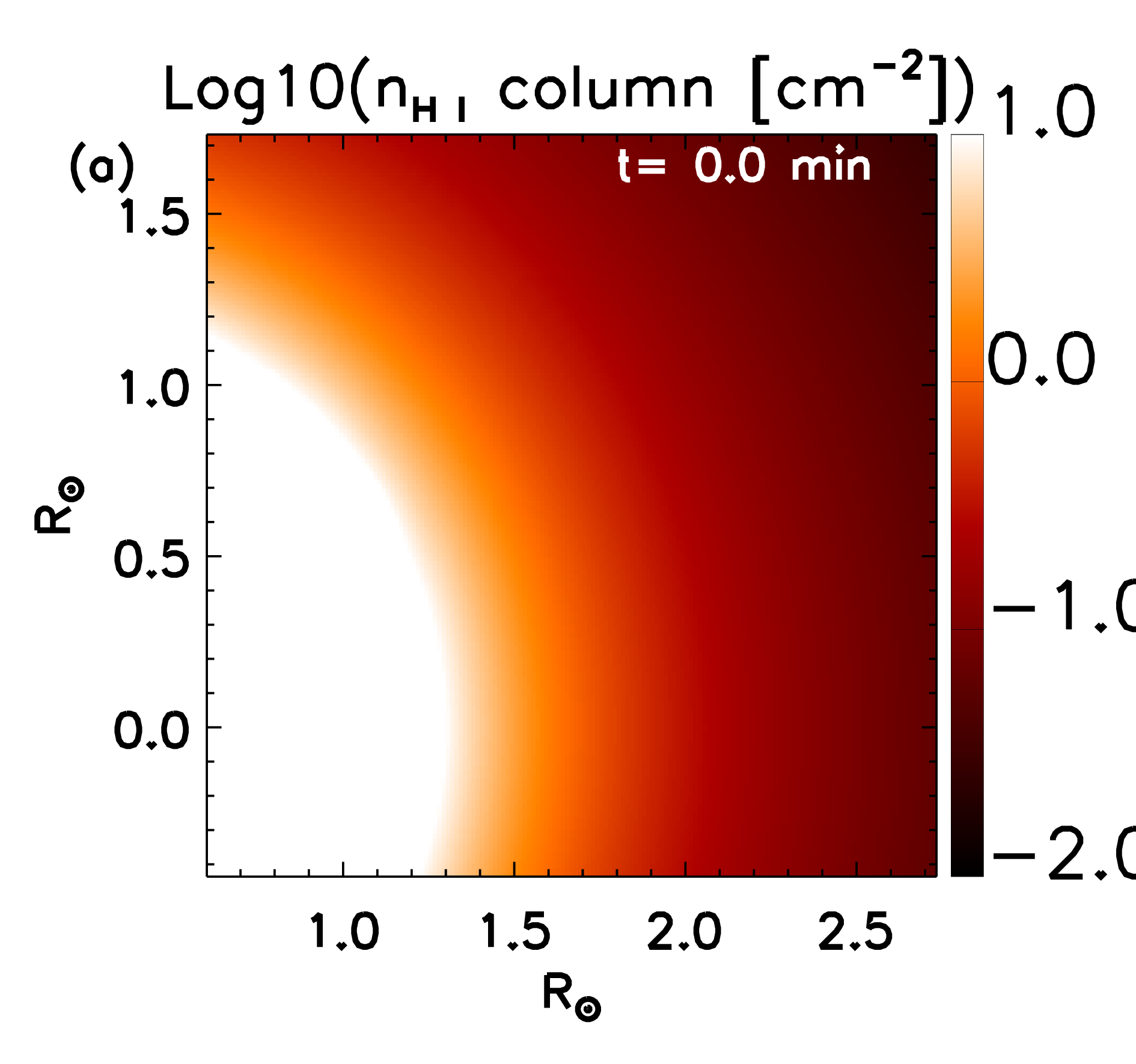}
\includegraphics[scale=0.13]{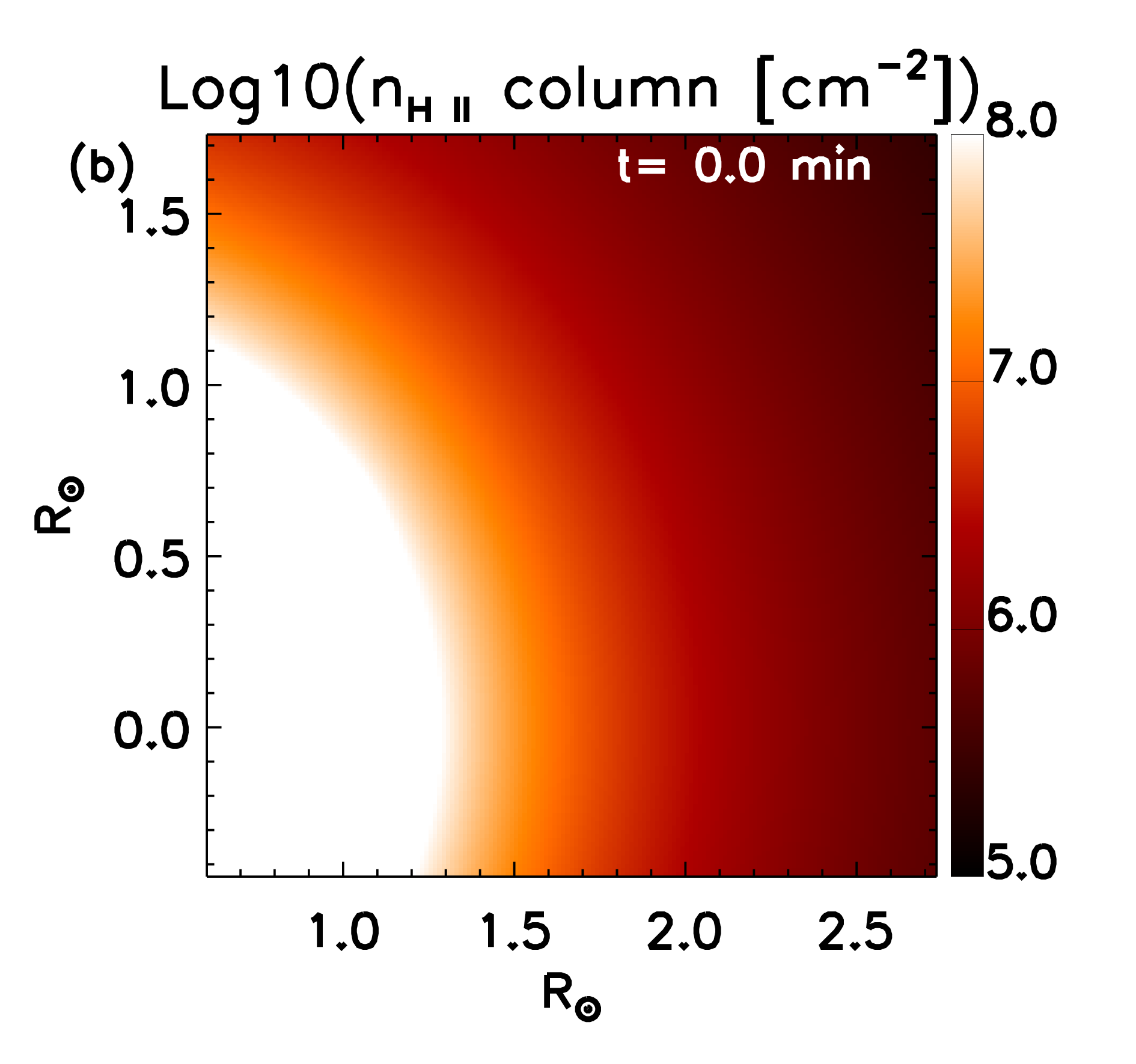}

\includegraphics[scale=0.13]{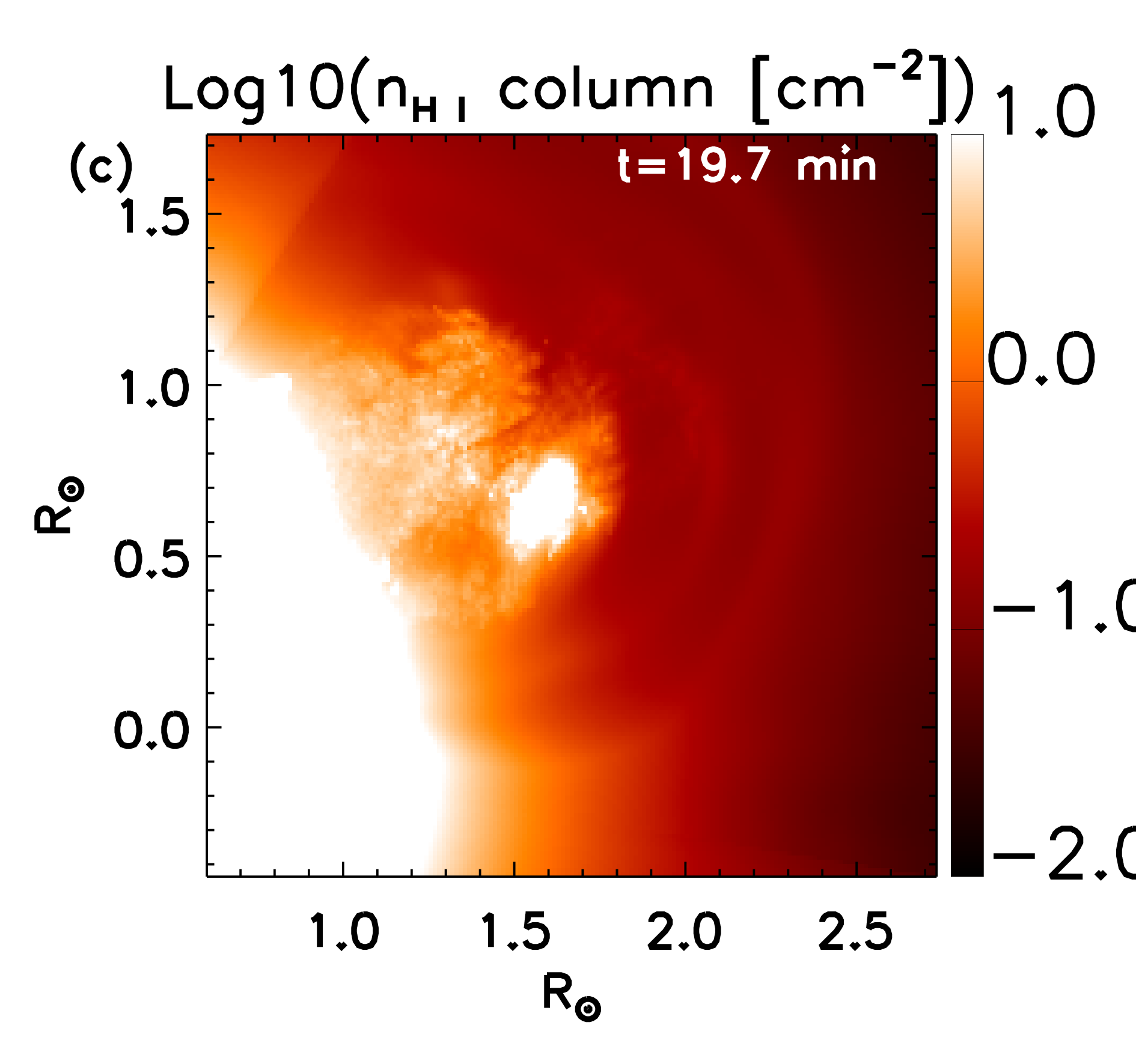}
\includegraphics[scale=0.13]{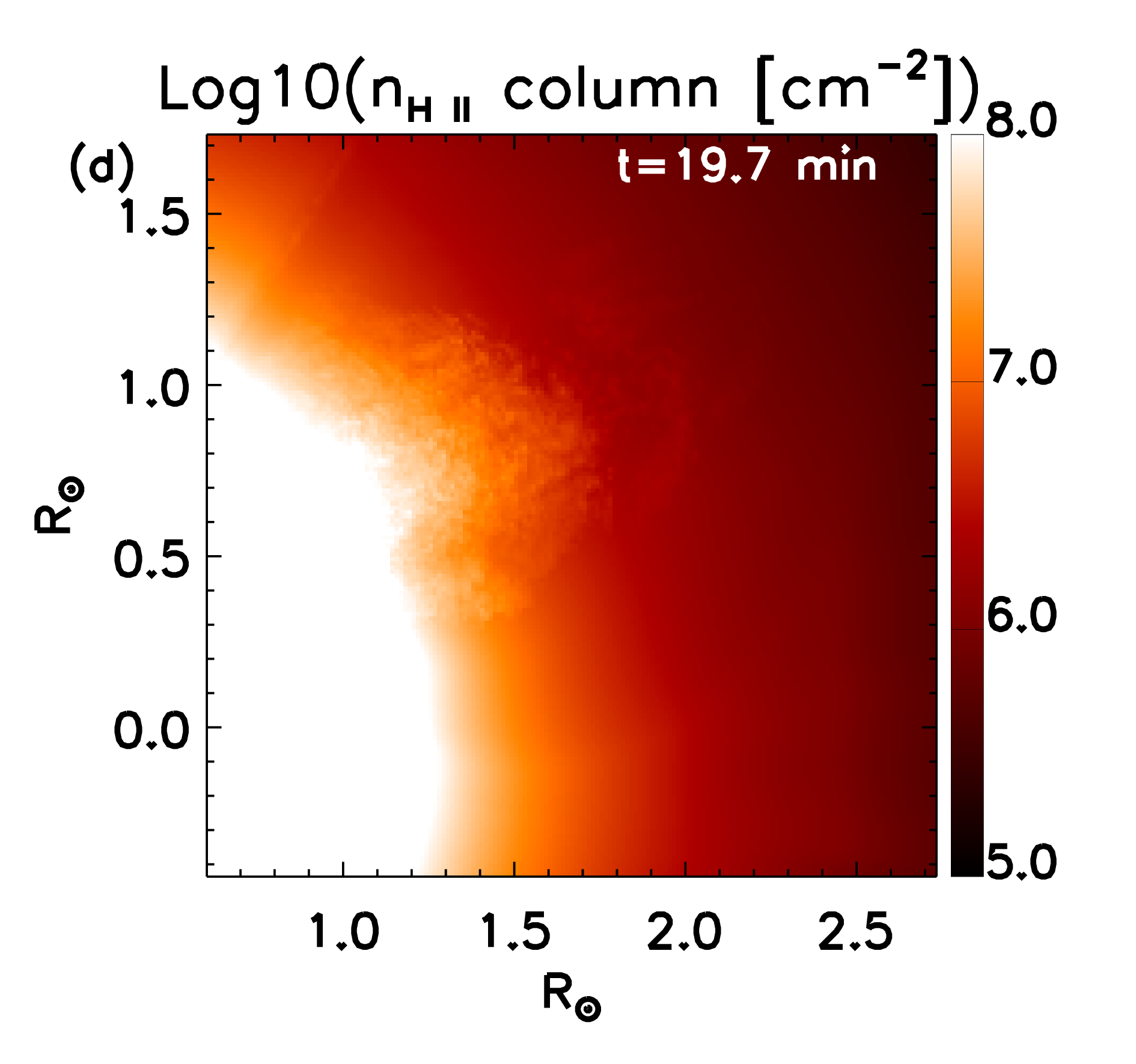}

\caption{Maps of the column density of neutral Hydrogen (a) and ionised Hydrogen (b) at $t=0$ $min$ and at $t=19.7$ $min$ - (c) and (d).}
\label{nhInhIImaps}
\end{figure}

Fig.\ref{nhInhIImaps} shows
the column density of $n_{HI}$ and $n_{HII}$
at $t=0$ and $t=19.7$ $min$
from the same point of view as in Fig.\ref{MHDneletemp}.
In order to display these results in terms of a potential observations we use a 3D cartesian box, where $x$ is the Sun East-West direction,
$y$ is the North-South direction, and $z$ is the direction along the LOS.
The evolution of both $n_{HI}$ and $n_{HII}$ follow a similar
pattern as in Fig.\ref{MHDneletemp}
where the distribution of $n_{HI}$ and $n_{HII}$
change significantly as soon as the magnetic flux rope ejection
propagates through the corona.
The CME three-parts structure is hinted in the distributions of 
$n_{HI}$ and $n_{HII}$.

\begin{figure*}
\centering
\includegraphics[scale=0.20]{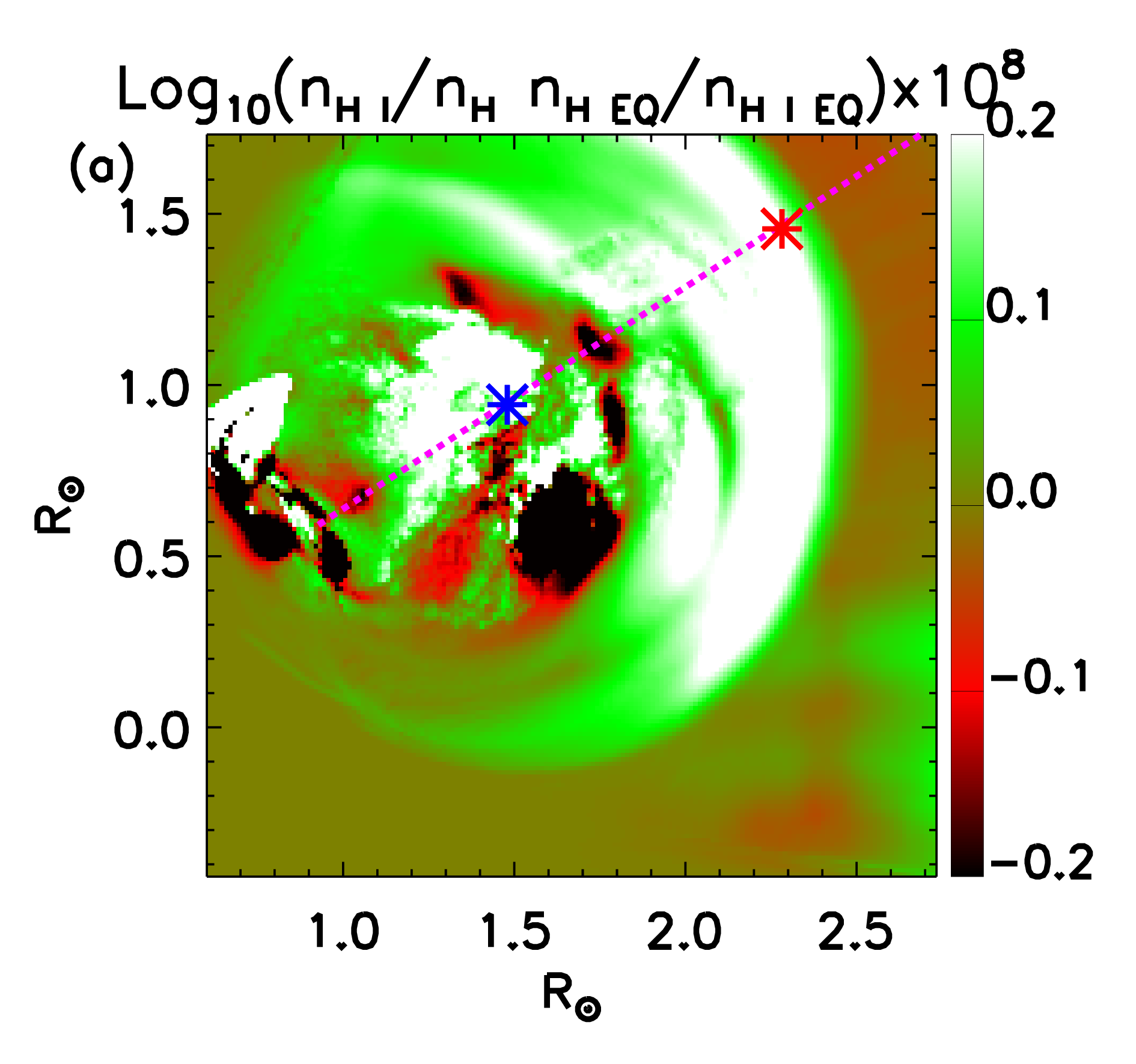}
\includegraphics[scale=0.20]{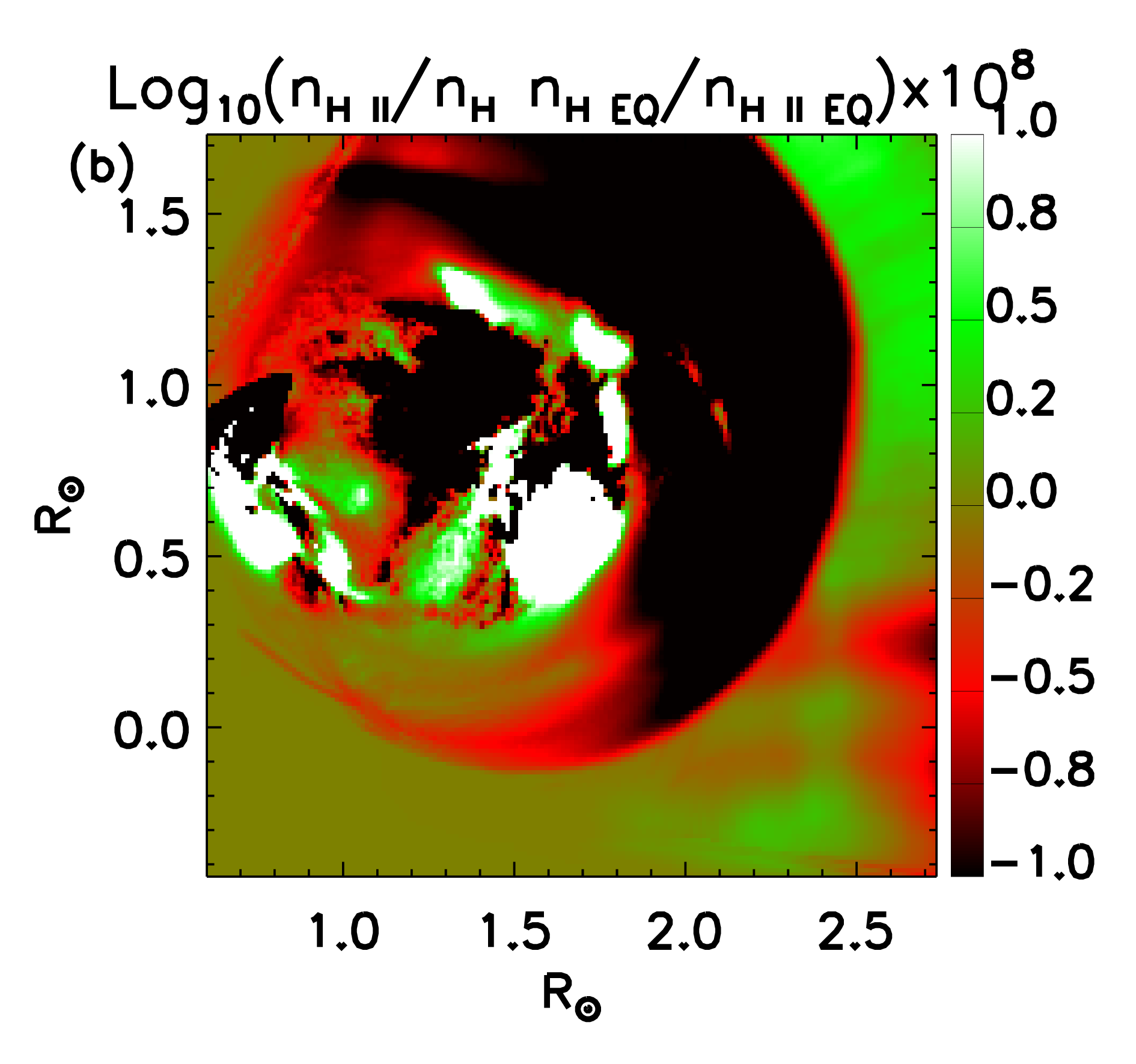}
\caption{Difference of the logarithms of the ionisation fraction for neutral (a) and ionised 
(b) Hydrogen atoms with respect to the values provided under the hypothesis of ionisation equilibrium at $t=19.7$ $min$.
The magenta line shows the trajectory of the CME on the plane of sky,
the blue and red asterisks denote the positions of the magnetic flux rope and of the CME front at this time, respectively.}
\label{nhInhIIoutmaps}
\end{figure*}

The key aim of this work is to 
measure the out-of-equilibrium ionisation effects 
in different CME components.
Therefore, in Fig.\ref{nhInhIIoutmaps} we show 
the difference of the logarithms of the ionisation fraction for the column densities of neutral and ionised Hydrogen at $t=19.7$ $min$ .
In these maps we find positive values when the non-equilibrium ionisation
effects lead to a cumulative higher number of neutral or ionised Hydrogen atoms 
compared to that of the ionisation equilibrium condition
along the LOS. Negative values occur when the number of neutral or ionised atoms is less than in the equilibrium condition.
We find extended regions where non-equilibrium ionisation effects become relevant.
This is particularly true for the abundance of neutral Hydrogen atoms where differences in logarithm between the 
number of atoms and the equilibrium condition along the LOS can be as large as 0.2, whereas
for the ionised Hydrogen atoms such differences are less than than $10^{-8}$
making non-equilibrium ionisation effects rather marginal.
In particular for the abundance of neutral Hydrogen we find 
that overall there are more atoms of this species along the LOS
than prescribed by ionisation equilibrium conditions and only in some small localised regions do we find less atoms of this species with respect to ionisation (Fig.\ref{nhInhIIoutmaps}a).

\begin{figure*}
\centering
\includegraphics[scale=0.60]{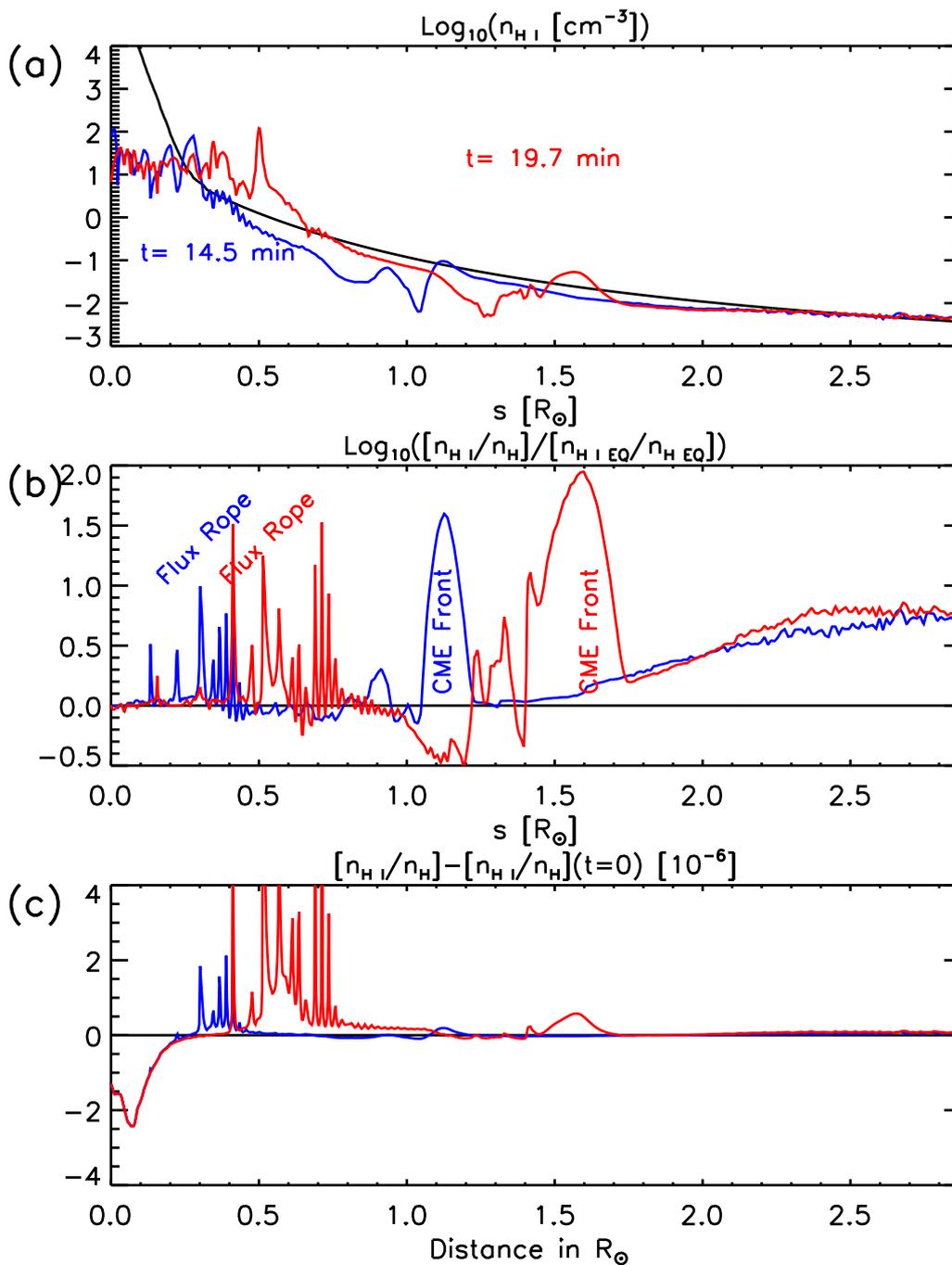}
\caption{(a) Radial profiles of neutral H atoms along the flux rope propagation direction
in the pre-CME corona (black line), and during the eruption at $t=14.5$ $min$ (blue line) and $t=19.7$ $min$ (red line).
(b) Comparison between ionisation equilibrium and non ionisation equilibrium cases at the same times, with labelled positions of the flux rope and CME front at different times.
(c) Changes in relative abundance of neutral Hydrogen at the same cuts and the same times.}
\label{nhinhiiradial}
\end{figure*}

As non-equilibrium ionisation effects are relevant for neutral Hydrogen in particular, 
we plot the number density of neutral Hydrogen in Fig.\ref{nhinhiiradial}a
at the initial condition, at $t=14.5$ $min$, and at $t=19.7$ $min$
along the direction of propagation of the magnetic flux rope.
If we focus on the curve at $t=19.7$ $min$
we find that the number of neutral Hydrogen atoms
increases with respect to the initial conditions 
where the magnetic flux rope is located ($s\sim0.7$ $R_{\odot}$),
as this is a structure colder than the ambient corona travelling upwards, 
thus carrying a higher abundance of 
neutral Hydrogen atoms through advection.
At the same time, the situation is more complex for the CME front ($s\sim1.6$ $R_{\odot}$)
where the number of neutral Hydrogen atoms has not changed significantly with respect to the 
pre-CME corona. The number of neutral Hydrogen atoms changes more because of the local temperature variations, i.e. ionisation and recombination processes rather than neutral Hydrogen atoms advection.

Comparing the neutral Hydrogen atoms abundance with the ionisation equilibrium conditions,
the difference is much more evident at the CME front than at the magnetic flux rope location (Fig.\ref{nhinhiiradial}b).
This is clearly visible in the cuts at $t=14.5$ $min$ and $t=19.7$ $min$.
At the magnetic flux rope location
the abundance of neutral Hydrogen atoms can be more than 10 times larger than 
those of ionisation equilibrium, but
it is also highly variable and this occurs in extremely localised regions.
In contrast, at the CME front the abundance of neutral Hydrogen atoms
is almost 100 times larger than that of ionisation equilibrium and this occurs consistently through the CME front for an extension of about $\sim0.3$ $R_{\odot}$.
In Fig.\ref{nhinhiiradial}c we show the change in the neutral Hydrogen relative abundance along the same direction, where we find an increase of neutral Hydrogen relative abundance with respect to the initial configuration at the flux rope location. As said, this is a natural consequence of the advection of the many neutral Hydrogen atoms from the initial position of the flux rope to higher altitude.
A small increase is found also at the CME front locations that is due to local
velocity fields near the front.

\begin{figure*}
\centering
\includegraphics[scale=0.60]{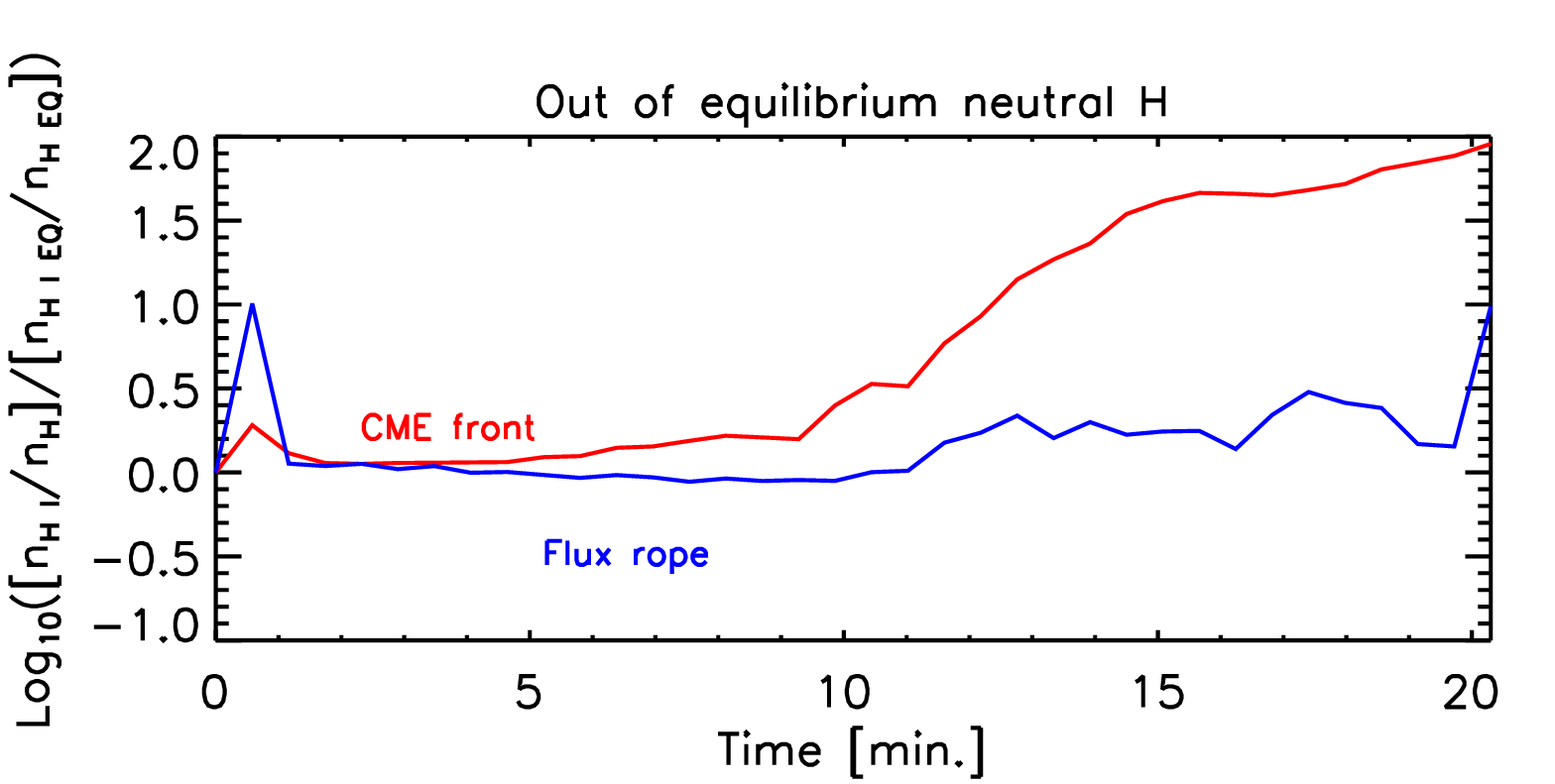}
\caption{Time evolution of relative abundance of neutral H atoms in the CME front (red line) and in the CME core (blue line); the curves show at each time the comparison between the ionisation 
equilibrium and non ionisation equilibrium cases.}
\label{nhinhiiouttime}
\end{figure*}

In Fig.\ref{nhinhiiouttime} we compare the neutral Hydrogen atoms abundance with the ionisation equilibrium conditions
for the centre of the magnetic flux rope location and the CME front as a function of time.
For the flux rope, the abundance of the neutral Hydrogen atoms remains very close to those of the one at ionisation equilibrium condition at all times, except for a very short lived peak at the beginning of the simulation when impulsive heating occurs. This shows that
the high density conditions in the flux rope allows for a quick settlement to ionisation equilibrium.
In contrast, the abundance of neutral Hydrogen atoms 
steadily departs from ionisation equilibrium conditions 
as soon as a CME front develops.
In our study we follow the ionisation state of Hydrogen atoms for approximately 20 minutes 
and at the end the abundance of the neutral Hydrogen atoms saturates
to a value that is about 100 times larger than the abundance expected from
ionisation equilibrium.

\begin{figure*}
\centering
\includegraphics[scale=0.60]{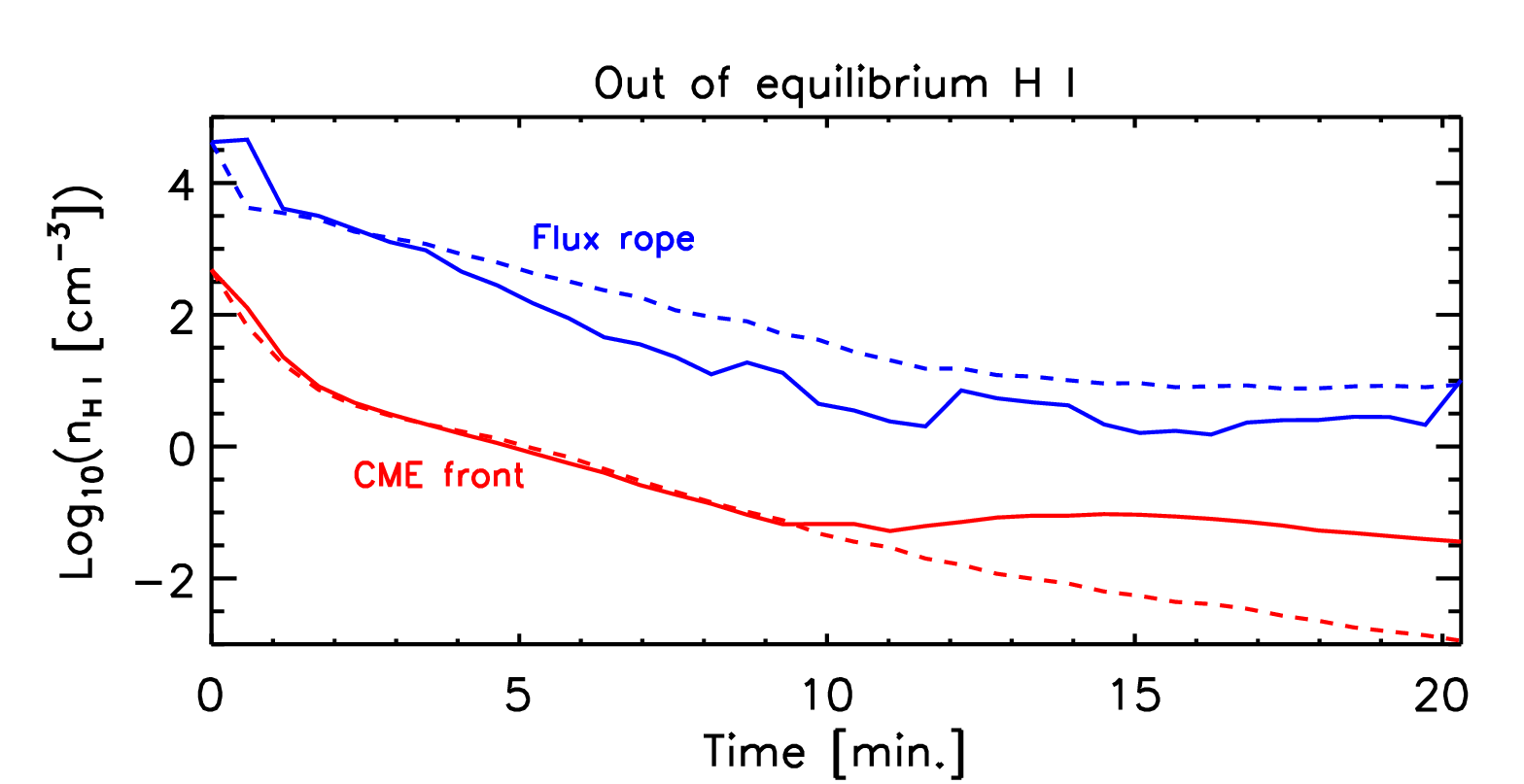}
\caption{Time evolution of absolute abundance of neutral Hydrogen atoms in the CME front (red line) and in the CME core (blue line), with respect to the ionisation equilibrium values (dashed lines).}
\label{nhinhiieqandouteq}
\end{figure*}

The physical explanation for this behaviour can be explained by comparing the
evolution of the number of neutral Hydrogen atoms in the magnetic flux rope and in the CME front 
as a function of time, which is shown in Fig.\ref{nhinhiieqandouteq}.
This is to be compared with Fig.\ref{frontfrtemprho} where we show the evolution of density and temperature in the same structures.
For the centre of the magnetic flux rope the expected density of neutral Hydrogen atoms at ionisation equilibrium 
steadily decreases (blue dashed curve) as the plasma density there decreases faster than the temperature does. 
While a lower temperature would lead to more neutral Hydrogen atoms, the overall lower density 
prescribes less particles in general. 
However, when we reconstruct the ionisation state of the plasma we find less neutral Hydrogen atoms in the 
magnetic flux rope (blue solid line) than at the ionisation equilibrium
as the ionisation state needs to adapt while the temperature is decreasing.
Also for the CME front, the expected density of neutral Hydrogen atoms at ionisation equilibrium
steadily decreases because of the steady decrease of the plasma density (red dashed curve), whereas
the CME front temperature remain more or less constant.
However, the decrease of neutral Hydrogen atoms is not as rapid (red solid curve),
because the front is initially loaded with a higher number of neutral Hydrogen atoms
and as the temperature increases, the ionisation state needs time to adapt to the changing temperature.

\begin{figure*}
\centering
\includegraphics[scale=0.60]{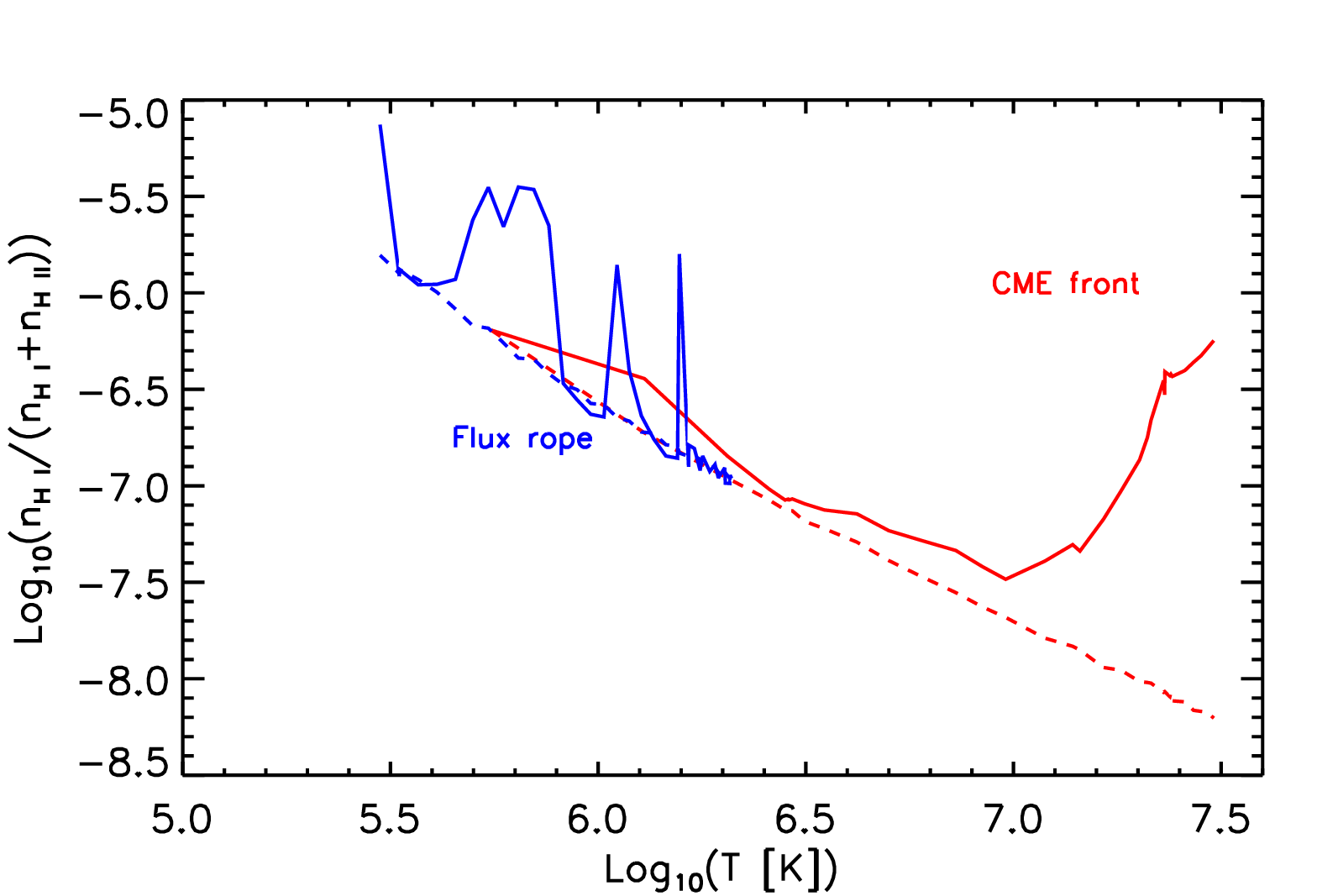}
\caption{Abundance of neutral Hydrogen as a function of temperature for the flux rope and CME front locations.}
\label{tempnhifrontfluxrope}
\end{figure*}

Fig.\ref{tempnhifrontfluxrope} shows the relative abundance of neutral Hydrogen atoms as a function of the temperature for the flux rope and the CME front.
This is compared with the abundance that would be measured at the ionisation equilibrium. Although this specific estimate is highly dependent on the CME model that is used here, 
it still bears some interesting results, as it happens that the flux rope is mostly near the ionisation equilibrium except in the range of $5.7<Log_{10}(T)<5.9$.
This is the temperature range in which the flux rope shortly after the eruption when the resistive heating occurs and it then converge to equilibrium as it cools down.
In contrast, the front remain close to ionisation equilibrium for lower temperatures $Log_{10}<6.9$ and then it significantly departs from equilibrium at higher temperatures. 

\begin{figure*}
\centering
\includegraphics[scale=0.45]{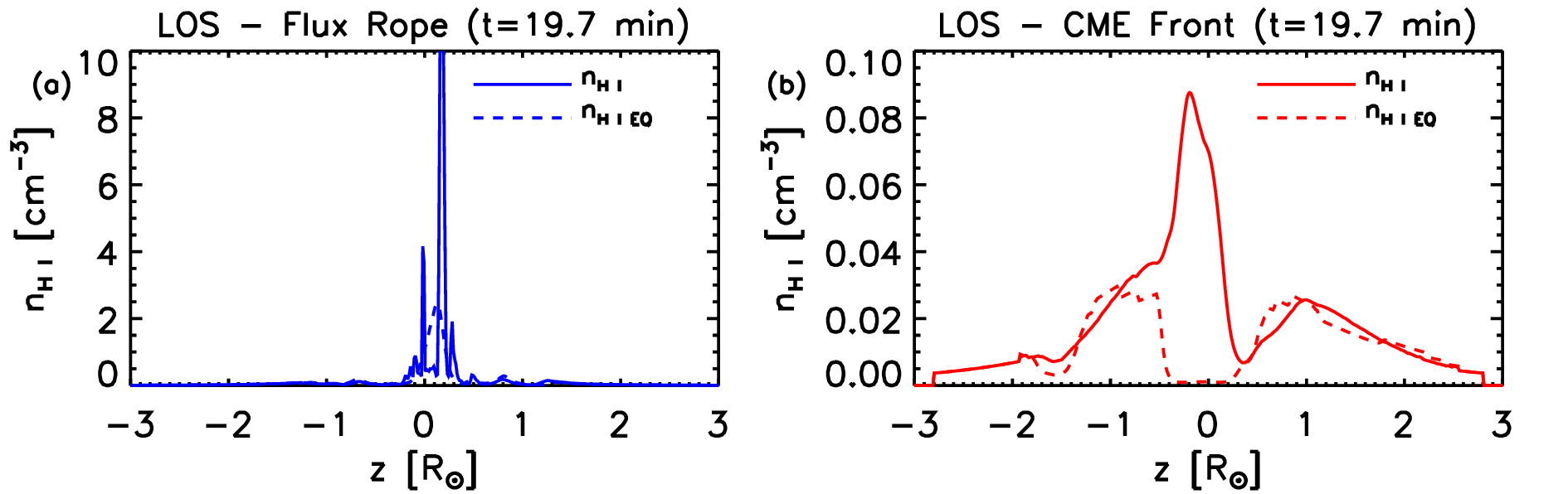}
\caption{(a) Neutral Hydrogen number density along the line of sight through the magnetic flux rope (solid line) and assuming ionisation equilibrium (dashed line) at $t=19.7$ $min$).
(b) Neutral Hydrogen number density along the line of sight through the CME front (solid line) and assuming ionisation equilibrium (dashed line) at $t=19.7$ $min$)}
\label{los_cuts_outeq}
\end{figure*}

It should be noted however that the unavoidable superposition of structures along the LOS can have a significant effect on these estimates.
Fig.\ref{los_cuts_outeq} shows the number density of neutral Hydrogen along the two lines of sight identified in Fig.\ref{nhInhIIoutmaps}, 
one cutting through the magnetic flux rope (Fig.\ref{los_cuts_outeq}a)
and the other through the CME front (Fig.\ref{los_cuts_outeq}b)
and it compares these densities with the values obtained assuming ionisation equilibrium.
The magnetic flux rope is a structure much denser than other structures along the LOS, therefore the measured ionisation state is not changing significantly whether we consider or not the integration along the LOS.
On the other hand, the CME front has a density comparable with the background density, hence the significantly larger amount of neutral Hydrogen atoms that we find at the front location is smoothed when the integration along the LOS is considered.
For this reason, the flux rope shows an abundance of neutral Hydrogen atoms that is on the order of 10 times (or less) the abundance in ionisation equilibrium and the same ratio is shown when we integrate along the LOS. The CME front, instead, shows an abundance of neutral Hydrogen atoms that is up to 100 times larger than the abundance in ionisation equilibrium, but when we integrate along the LOS the count of neutral Hydrogen atoms is only about 10 times larger than the ionisation equilibrium conditions.

\section{Discussion and Conclusions}
\label{conclusions}

In this work we have used the MHD simulation of a magnetic flux rope ejection and CME propagation 
in the solar corona from \citet{Pagano2014} to study the abundances of ionised and neutral Hydrogen atoms 
during CMEs.
In particular, we focus on the non-ionisation equilibrium effects in order to verify to what extend the assumption of
ionisation equilibrium holds during these phenomena.
This work is especially important for the forthcoming measurements of Metis, the coronagraph on board the Solar Orbiter mission, and LST, the coronagraph on-board ASO-S,
that will observe the solar corona in the Lyman-$\alpha$ line that is produced by neutral Hydrogen atoms.
A number of measurements of plasma properties, such as the plasma temperature, crucially depend 
on the amount of neutral Hydrogen atoms and on the assumptions that are been made for the ionisation state of the plasma.
For example, one way to derive the plasma temperature is to compare the abundances of neutral and ionised Hydrogen atoms
and to find the temperature at which the measured abundances are in agreement with the theoretical ones estimated by the ionisation equilibrium hypothesis.
Such kinds of derivations are inherently prone to errors if the plasma is not in ionisation equilibrium.
To this end, this work sheds light on the regions and phenomena  that occur when the assumption of ionisation equilibrium holds and when it does not.

In order to reconstruct the ionisation state of Hydrogen atoms during a CME propagation in the solar corona,
we use the MHD simulation of \citet{Pagano2014} that provides the 3D plasma density and temperature distributions 
as a function of time, together with the plasma velocity fields.
Using these as inputs and assuming ionisation equilibrium at the beginning of the CME propagation,
we devised a technique to solve the differential equations that describe the change in the number of neutral and ionised Hydrogen atoms as a function of space and time, using the strategy already introduced in \citet{Pagano2008}.
More specifically, we separately estimate the advection and ionisation/recombination terms to 
express the rate of change of the number of ionised and neutral Hydrogen atoms in one cell of the MHD simulation.
By time-integrating the rate of change we can describe abundance of neutral and ionised Hydrogen atoms
as a function of space and time in our MHD simulation.
This technique bears no assumption on the ionisation state of the plasma 
and it can describe abundances of ionised and neutral Hydrogen atoms out of ionisation equilibrium.

Finally, to analyse the results of this study with regards to the temperature measurements in the solar corona,
we have compared the obtained distributions of neutral and ionised Hydrogen atoms
with respect to the distributions associated with ionisation equilibrium.
We do this through  the entire 3D domain of the MHD simulation and specifically for 
the centre of the magnetic flux rope and the CME front
that are two prominent features of CMEs.

We find that the departures from ionisation equilibrium are negligible for the abundance of ionised Hydrogen atoms, where any
departure is small enough to become irrelevant, as the vast majority of Hydrogen atoms are ionised in the solar corona.
However, this is not the case for  neutral Hydrogen since there are inherently few 
numbers of these atoms in the corona 
and therefore small variations can lead to a significant departure from the assumed abundance at ionisation equilibrium.
We crucially find that non-equilibrium ionisation effects for neutral Hydrogen atoms are marginal for the regions pertinent to the magnetic flux rope.
This means that the amount of neutral Hydrogen atoms in the magnetic flux rope is generally consistent with the number 
one would assume in ionisation equilibrium conditions.
Although we observe some small scale variations and rapid changes from one location to another within the magnetic flux rope,
overall we find that these effects are not detectable when the whole flux rope is considered.
This means that, for the future derivation of plasma temperatures based on observed Lyman-$\alpha$ line
intensities, the ionisation equilibrium hypothesis is still applicable.

In contrast, we find that non-equilibrium ionisation effects are relevant at the CME front,
where the number of neutral Hydrogen is much larger than the expected abundance from the assumption of ionisation equilibrium conditions.
We find this to be consistent over the whole extension of the CME front and that
the departure is as large as 2 orders of magnitude in excess with respect to the ionisation equilibrium conditions. In the region where we have an excess of neutral Hydrogen atoms, the derived temperature from the abundances would be much lower than 
the actual temperature value, and thus using the ionisation equilibrium assumption would lead to a significant plasma temperature underestimate at the CME front.
At the same time, we also find that the projection effect along the LOS can partially hide these effects as it happens here for the CME front where non-equilbrium ionisation effects are smoothed out after the integration along the LOS and the superposition of structures in ionisation equilibirium that have comparable densities.

In the future more studies need to be carried out, to investigate the parameter space of the CME velocities, densities and temperatures, along with considering more diverse configurations of the magnetic flux rope ejections. It is also important to apply this approach (the magnetofrictional simulation, the MHD simulation, and the reconstruction of the ionisation state of the Hydrogen atoms) 
to observed events to more accurately verify these results.
In particular, the analysis carried out in Fig.\ref{tempnhifrontfluxrope} can potentially be extended to many more cases where MHD simulations can be used to derive a more accurate and reliable lookup table to invert the temperature of CME fronts and cores as a function of the relative abundance of neutral Hydrogen.
At the same time, this work already provides key information for a correct interpretation of future plasma properties derived from Lyman-$\alpha$ observations, which is
certainly crucial for the future observations of Metis on board Solar Orbiter and LST on board ASO-S telescopes.

\begin{acknowledgements}

This research has received funding from the Science and Technology Facilities Council (UK) through the consolidated grant ST/N000609/1 and the European Research Council (ERC) under the European Union Horizon 2020 research and innovation program (grant agreement No. 647214).
D.H.M. would like to thank both the UK STFC and the ERC (Synergy grant: WHOLE SUN, grant Agreement No. 810218) for financial support. D.H.M. and P.P. would like to thank STFC for IAA funding under grant number SMC1-XAS012.
This work used the DiRAC@Durham facility managed by the Institute for Computational Cosmology on behalf of the STFC DiRAC HPC Facility (www.dirac.ac.uk. The equipment was funded by BEIS capital funding via STFC capital grants ST/P002293/1, ST/R002371/1 and ST/S002502/1, Durham University and STFC operations grant ST/R000832/1. DiRAC is part of the National e-Infrastructure.
We acknowledge the use of the open source (gitorious.org/amrvac) MPI-AMRVAC software, relying on coding efforts from C. Xia, O. Porth, R. Keppens.
CHIANTI is a collaborative project involving George Mason University, the University of Michigan (USA), University of Cambridge (UK) and NASA Goddard Space Flight Center (USA).

\end{acknowledgements}

\bibliographystyle{aa}
\bibliography{ref}

\begin{thebibliography}{54}
\expandafter\ifx\csname natexlab\endcsname\relax\def\natexlab#1{#1}\fi

\bibitem[{{Akmal} {et~al.}(2001){Akmal}, {Raymond}, {Vourlidas}, {Thompson},
  {Ciaravella}, {Ko}, {Uzzo}, \& {Wu}}]{akmal2001}
{Akmal}, A., {Raymond}, J.~C., {Vourlidas}, A., {et~al.} 2001, \apj, 553, 922

\bibitem[{{Antonucci} {et~al.}(2017){Antonucci}, {Andretta}, {Cesare},
  {Ciaravella}, {Doschek}, {Fineschi}, {Giordano}, {Lamy}, {Moses}, {Naletto},
  {Newmark}, {Poletto}, {Romoli}, {Solanki}, {Spadaro}, {Teriaca}, \&
  {Zangrilli}}]{2017SPIE10566E..0LA}
{Antonucci}, E., {Andretta}, V., {Cesare}, S., {et~al.} 2017, in Society of
  Photo-Optical Instrumentation Engineers (SPIE) Conference Series, Vol. 10566,
  \procspie, 105660L

\bibitem[{{Aschwanden}(2017)}]{aschwanden2017}
{Aschwanden}, M.~J. 2017, \apj, 847, 27

\bibitem[{{Bemporad} \& {Pagano}(2015)}]{BemporadPagano2015}
{Bemporad}, A. \& {Pagano}, P. 2015, \aap, 576, A93

\bibitem[{{Bemporad} {et~al.}(2018){Bemporad}, {Pagano}, \&
  {Giordano}}]{Bemporad2018}
{Bemporad}, A., {Pagano}, P., \& {Giordano}, S. 2018, \aap, 619, A25

\bibitem[{{Bradshaw}(2009)}]{bradshaw2009}
{Bradshaw}, S.~J. 2009, \aap, 502, 409

\bibitem[{{Bradshaw} \& {Testa}(2019)}]{2019ApJ...872..123B}
{Bradshaw}, S.~J. \& {Testa}, P. 2019, \apj, 872, 123

\bibitem[{{Chen}(2011)}]{Chen2011}
{Chen}, P.~F. 2011, Living Reviews in Solar Physics, 8, 1

\bibitem[{{Cheng} {et~al.}(2010){Cheng}, {Ding}, \& {Zhang}}]{Cheng2010}
{Cheng}, X., {Ding}, M.~D., \& {Zhang}, J. 2010, \apj, 712, 1302

\bibitem[{{Ciaravella} {et~al.}(2006){Ciaravella}, {Raymond}, \&
  {Kahler}}]{Ciaravella2006}
{Ciaravella}, A., {Raymond}, J.~C., \& {Kahler}, S.~W. 2006, \apj, 652, 774

\bibitem[{{Ciaravella} {et~al.}(2001){Ciaravella}, {Raymond}, {Reale},
  {Strachan}, \& {Peres}}]{Ciaravella2001}
{Ciaravella}, A., {Raymond}, J.~C., {Reale}, F., {Strachan}, L., \& {Peres}, G.
  2001, \apj, 557, 351

\bibitem[{{Colgan} {et~al.}(2008){Colgan}, {Abdallah}, {Sherrill}, {Foster},
  {Fontes}, \& {Feldman}}]{Colgan2008}
{Colgan}, J., {Abdallah}, J., J., {Sherrill}, M.~E., {et~al.} 2008, \apj, 689,
  585

\bibitem[{{Courant} {et~al.}(1928){Courant}, {Friedrichs}, \&
  {Lewy}}]{1928MatAn.100...32C}
{Courant}, R., {Friedrichs}, K., \& {Lewy}, H. 1928, Mathematische Annalen,
  100, 32

\bibitem[{{Dere} {et~al.}(2019){Dere}, {Del Zanna}, {Young}, {Landi}, \&
  {Sutherland}}]{dere2019}
{Dere}, K.~P., {Del Zanna}, G., {Young}, P.~R., {Landi}, E., \& {Sutherland},
  R.~S. 2019, \apjs, 241, 22

\bibitem[{{Dud{\'\i}k} {et~al.}(2017){Dud{\'\i}k}, {Dzif{\v{c}}{\'a}kov{\'a}},
  {Meyer-Vernet}, {Del Zanna}, {Young}, {Giunta}, {Sylwester}, {Sylwester},
  {Oka}, {Mason}, {Vocks}, {Matteini}, {Krucker}, {Williams}, \&
  {Mackovjak}}]{dudik2017}
{Dud{\'\i}k}, J., {Dzif{\v{c}}{\'a}kov{\'a}}, E., {Meyer-Vernet}, N., {et~al.}
  2017, \solphys, 292, 100

\bibitem[{{Dud{\'\i}k} {et~al.}(2014){Dud{\'\i}k}, {Janvier}, {Aulanier}, {Del
  Zanna}, {Karlick{\'y}}, {Mason}, \& {Schmieder}}]{dudik2014}
{Dud{\'\i}k}, J., {Janvier}, M., {Aulanier}, G., {et~al.} 2014, \apj, 784, 144

\bibitem[{{Frassati} {et~al.}(2019){Frassati}, {Susino}, {Mancuso}, \&
  {Bemporad}}]{frassati2019}
{Frassati}, F., {Susino}, R., {Mancuso}, S., \& {Bemporad}, A. 2019, \apj, 871,
  212

\bibitem[{{Godunov}(1959)}]{Godunov1959}
{Godunov}, S. 1959, Math. Sbornik, 47, 271

\bibitem[{{Hannah} \& {Kontar}(2013)}]{hannah2013}
{Hannah}, I.~G. \& {Kontar}, E.~P. 2013, \aap, 553, A10

\bibitem[{{Howard} \& {DeForest}(2014)}]{HowardDeForest2014}
{Howard}, T.~A. \& {DeForest}, C.~E. 2014, \apj, 796, 33

\bibitem[{{Imada} {et~al.}(2011){Imada}, {Hara}, {Watanabe}, {Murakami},
  {Harra}, {Shimizu}, \& {Zweibel}}]{imada2011}
{Imada}, S., {Hara}, H., {Watanabe}, T., {et~al.} 2011, \apj, 743, 57

\bibitem[{{Jej{\v{c}}i{\v{c}}} {et~al.}(2017){Jej{\v{c}}i{\v{c}}}, {Susino},
  {Heinzel}, {Dzif{\v{c}}{\'a}kov{\'a}}, {Bemporad}, \& {Anzer}}]{jejcic2017}
{Jej{\v{c}}i{\v{c}}}, S., {Susino}, R., {Heinzel}, P., {et~al.} 2017, \aap,
  607, A80

\bibitem[{{Karlick{\'y}} {et~al.}(2004){Karlick{\'y}}, {Ka{\v{s}}parov{\'a}},
  \& {Heinzel}}]{2004A&A...416L..13K}
{Karlick{\'y}}, M., {Ka{\v{s}}parov{\'a}}, J., \& {Heinzel}, P. 2004, \aap,
  416, L13

\bibitem[{{Kocher} {et~al.}(2018){Kocher}, {Landi}, \& {Lepri}}]{kocher2018}
{Kocher}, M., {Landi}, E., \& {Lepri}, S.~T. 2018, \apj, 860, 51

\bibitem[{{Landi} {et~al.}(2010){Landi}, {Raymond}, {Miralles}, \&
  {Hara}}]{landi2010}
{Landi}, E., {Raymond}, J.~C., {Miralles}, M.~P., \& {Hara}, H. 2010, \apj,
  711, 75

\bibitem[{{Lee} {et~al.}(2019{\natexlab{a}}){Lee}, {Raymond}, {Reeves}, {Shen},
  {Moon}, \& {Kim}}]{2019ApJ...879..111L}
{Lee}, J.-Y., {Raymond}, J.~C., {Reeves}, K.~K., {et~al.} 2019{\natexlab{a}},
  \apj, 879, 111

\bibitem[{{Lee} {et~al.}(2019{\natexlab{b}}){Lee}, {Raymond}, {Reeves}, {Shen},
  {Moon}, \& {Kim}}]{lee2019}
{Lee}, J.-Y., {Raymond}, J.~C., {Reeves}, K.~K., {et~al.} 2019{\natexlab{b}},
  \apj, 879, 111

\bibitem[{{Li}(2016)}]{2016IAUS..320..436L}
{Li}, H. 2016, in IAU Symposium, Vol. 320, Solar and Stellar Flares and their
  Effects on Planets, ed. A.~G. {Kosovichev}, S.~L. {Hawley}, \& P.~{Heinzel},
  436--438

\bibitem[{{Mackay} {et~al.}(2011){Mackay}, {Green}, \& {van
  Ballegooijen}}]{Mackay2011}
{Mackay}, D.~H., {Green}, L.~M., \& {van Ballegooijen}, A. 2011, \apj, 729, 97

\bibitem[{{Mackay} \& {van Ballegooijen}(2006)}]{MackayVanBallegooijen2006A}
{Mackay}, D.~H. \& {van Ballegooijen}, A.~A. 2006, \apj, 641, 577

\bibitem[{{Mart{\'\i}nez-Sykora} {et~al.}(2016){Mart{\'\i}nez-Sykora}, {De
  Pontieu}, {Hansteen}, \& {Gudiksen}}]{2016ApJ...817...46M}
{Mart{\'\i}nez-Sykora}, J., {De Pontieu}, B., {Hansteen}, V.~H., \& {Gudiksen},
  B. 2016, \apj, 817, 46

\bibitem[{{Murphy} {et~al.}(2011){Murphy}, {Raymond}, \&
  {Korreck}}]{murphy2011}
{Murphy}, N.~A., {Raymond}, J.~C., \& {Korreck}, K.~E. 2011, \apj, 735, 17

\bibitem[{{Noci} {et~al.}(1987){Noci}, {Kohl}, \& {Withbroe}}]{Noci1987}
{Noci}, G., {Kohl}, J.~L., \& {Withbroe}, G.~L. 1987, \apj, 315, 706

\bibitem[{{Pagano} {et~al.}(2015{\natexlab{a}}){Pagano}, {Bemporad}, \&
  {Mackay}}]{Pagano2015b}
{Pagano}, P., {Bemporad}, A., \& {Mackay}, D.~H. 2015{\natexlab{a}}, \aap, 582,
  A72

\bibitem[{{Pagano} {et~al.}(2013{\natexlab{a}}){Pagano}, {Mackay}, \&
  {Poedts}}]{Pagano2013b}
{Pagano}, P., {Mackay}, D.~H., \& {Poedts}, S. 2013{\natexlab{a}}, \aap, 560,
  A38

\bibitem[{{Pagano} {et~al.}(2013{\natexlab{b}}){Pagano}, {Mackay}, \&
  {Poedts}}]{Pagano2013a}
{Pagano}, P., {Mackay}, D.~H., \& {Poedts}, S. 2013{\natexlab{b}}, \aap, 554,
  A77

\bibitem[{{Pagano} {et~al.}(2014){Pagano}, {Mackay}, \& {Poedts}}]{Pagano2014}
{Pagano}, P., {Mackay}, D.~H., \& {Poedts}, S. 2014, \aap, 568, A120

\bibitem[{{Pagano} {et~al.}(2015{\natexlab{b}}){Pagano}, {Mackay}, \&
  {Poedts}}]{Pagano2015a}
{Pagano}, P., {Mackay}, D.~H., \& {Poedts}, S. 2015{\natexlab{b}}, Journal of
  Astrophysics and Astronomy, 36, 123

\bibitem[{{Pagano} {et~al.}(2018){Pagano}, {Pascoe}, \& {De
  Moortel}}]{Pagano2018}
{Pagano}, P., {Pascoe}, D., \& {De Moortel}, I. 2018, \aap, 999, A999

\bibitem[{{Pagano} {et~al.}(2008){Pagano}, {Raymond}, {Reale}, \&
  {Orlando}}]{Pagano2008}
{Pagano}, P., {Raymond}, J.~C., {Reale}, F., \& {Orlando}, S. 2008, \aap, 481,
  835

\bibitem[{{Porth} {et~al.}(2014){Porth}, {Xia}, {Hendrix}, {Moschou}, \&
  {Keppens}}]{Porth2014}
{Porth}, O., {Xia}, C., {Hendrix}, T., {Moschou}, S.~P., \& {Keppens}, R. 2014,
  \apjs, 214, 4

\bibitem[{{Raymond} {et~al.}(2003){Raymond}, {Ghavamian}, {Sankrit}, {Blair},
  \& {Curiel}}]{Raymond2003a}
{Raymond}, J.~C., {Ghavamian}, P., {Sankrit}, R., {Blair}, W.~P., \& {Curiel},
  S. 2003, \apj, 584, 770

\bibitem[{{Reep} {et~al.}(2019){Reep}, {Bradshaw}, {Crump}, \&
  {Warren}}]{2019ApJ...871...18R}
{Reep}, J.~W., {Bradshaw}, S.~J., {Crump}, N.~A., \& {Warren}, H.~P. 2019,
  \apj, 871, 18

\bibitem[{{Rodkin} {et~al.}(2017){Rodkin}, {Goryaev}, {Pagano}, {Gibb},
  {Slemzin}, {Shugay}, {Veselovsky}, \& {Mackay}}]{Rodkin2017}
{Rodkin}, D., {Goryaev}, F., {Pagano}, P., {et~al.} 2017, \solphys, 292, 90

\bibitem[{{Shen} {et~al.}(2017){Shen}, {Raymond}, {Miki{\'c}}, {Linker},
  {Reeves}, \& {Murphy}}]{2017ApJ...850...26S}
{Shen}, C., {Raymond}, J.~C., {Miki{\'c}}, Z., {et~al.} 2017, \apj, 850, 26

\bibitem[{{Shen} {et~al.}(2013){Shen}, {Reeves}, {Raymond}, {Murphy}, {Ko},
  {Lin}, {Miki{\'c}}, \& {Linker}}]{shen2013}
{Shen}, C., {Reeves}, K.~K., {Raymond}, J.~C., {et~al.} 2013, \apj, 773, 110

\bibitem[{{Shi} {et~al.}(2019){Shi}, {Landi}, \& {Manchester}}]{shi2019}
{Shi}, T., {Landi}, E., \& {Manchester}, Ward, I. 2019, \apj, 882, 154

\bibitem[{{Smith} \& {Hughes}(2010)}]{2010ApJ...718..583S}
{Smith}, R.~K. \& {Hughes}, J.~P. 2010, \apj, 718, 583

\bibitem[{{Song} {et~al.}(2014){Song}, {Zhang}, {Chen}, \& {Cheng}}]{Song2014}
{Song}, H.~Q., {Zhang}, J., {Chen}, Y., \& {Cheng}, X. 2014, \apjl, 792, L40

\bibitem[{{Spitzer}(1962)}]{Spitzer1962}
{Spitzer}, L. 1962, {Physics of Fully Ionized Gases} (Physics of Fully Ionized
  Gases, New York: Interscience (2nd edition), 1962)

\bibitem[{{Su} {et~al.}(2016){Su}, {Cheng}, {Ding}, {Chen}, {Ning}, \&
  {Ji}}]{su2016}
{Su}, W., {Cheng}, X., {Ding}, M.~D., {et~al.} 2016, \apj, 830, 70

\bibitem[{{Susino} \& {Bemporad}(2016)}]{susino2016}
{Susino}, R. \& {Bemporad}, A. 2016, \apj, 830, 58

\bibitem[{{Yan} {et~al.}(2017){Yan}, {Jiang}, {Xue}, {Wang}, {Priest}, {Yang},
  {Kong}, {Cao}, \& {Ji}}]{Yan2017}
{Yan}, X.~L., {Jiang}, C.~W., {Xue}, Z.~K., {et~al.} 2017, \apj, 845, 18

\bibitem[{{Ying} {et~al.}(2019){Ying}, {Bemporad}, {Giordano}, {Pagano},
  {Feng}, {Lu}, {Li}, \& {Gan}}]{2019ApJ...880...41Y}
{Ying}, B., {Bemporad}, A., {Giordano}, S., {et~al.} 2019, \apj, 880, 41

\end{thebibliography}

\end{document}